\documentclass[12pt,psf,epsf]{article}
\usepackage[utf8]{inputenc}
\usepackage[english]{babel}

\usepackage{amsmath}
\usepackage{amssymb}
\usepackage{mathrsfs}
\usepackage{amsfonts}
\usepackage{amsthm}
\usepackage{mathtools}
\usepackage{jheppub}

\usepackage{graphicx}
\usepackage{epsfig}
\usepackage{subfig}

\usepackage{array}
\usepackage{latexsym}
\usepackage[mathcal]{euscript}
\usepackage{hyperref}

\usepackage{psfrag}
\usepackage{float}
\usepackage{cancel}
\usepackage{slashed}
\usepackage{bm}
\usepackage{color}
\usepackage{setspace}
\usepackage{color}

\usepackage{etoolbox}

\newcommand{\bea}{\begin{eqnarray}}
\newcommand{\eea}{\end{eqnarray}}

\definecolor{aquamarine}{rgb}{0.8,0.0,0.8}

\begin{document}

\title{Higgs-dilaton model revisited: can dilaton act as QCD axion?}

\author[a,b]{Aleksandr I. Belokon,}
\author[c]{Anna Tokareva}

\affiliation[a]{Steklov Mathematical Institute, Russian Academy of Sciences, Gubkin str. 8, 119991 Moscow, Russian Federation}

\affiliation[b]{Institute for Nuclear Research, Russian Academy of Sciences, Prospekt 60-letiya Oktyabrya 7a, 117312 Moscow, Russian Federation}

\affiliation[c]{Theoretical Physics, Blackett Laboratory, Imperial College London, SW7 2AZ London, United Kingdom}

\emailAdd{belokon@mi-ras.ru}
\emailAdd{a.tokareva@imperial.ac.uk}
\preprint{Imperial/TP/2022/AAT/3}
\abstract{The Standard Model Lagrangian has an approximate scale symmetry in the high-energy limit. This observation can be embedded into the fundamental principle of the ultimate theory of Nature as the requirement of the exact quantum scale invariance. In this setup, all low-energy particle phenomenology can be obtained as a result of the spontaneous breaking of scale symmetry leading to the presence of the extra massless scalar field --- the dilaton. We explore the scenario, in which this field is also capable of solving the strong CP-problem in QCD in a similar way as in QCD axion models. The dilaton, playing the role of the axion, is coupled to the QCD sector and acquires a mass term due to non-perturbative breaking of scale symmetry. We show that the dilaton can form dark matter after the low-scale inflation. As a proof of concept, we construct a model of Higgs-driven inflation consistently realising the axion-like dilaton dark matter production compatible with the CMB data.}

\maketitle

\newpage


\section{Introduction}

The Standard Model (SM) is a theory of low energy physics that perfectly explains the majority of experimental data. At energies much larger than the electroweak scale, it has a non-trivial property of classical scale symmetry. This looks as a hint that the complete theory might be scale-invariant at high energies. In particular, the idea that the scale symmetry, being restored at high energies, is spontaneously broken in the low energy domain looks attractive, because it allows to explain the absence of quadratic divergences and decrease the radiative corrections to the Higgs mass and vacuum energy~\cite{WETTERICH1984215,Bardeen:1995kv,Shaposhnikov:2018xkv,Shaposhnikov:2018jag,Shaposhnikov:2020geh}. However, it is hard to save scale invariance when constructing a quantum theory, since this symmetry is known to be anomalous. 

The solution to this problem was first suggested in~\cite{Englert:1976ep} and then developed in~\cite{Shaposhnikov:2008xb,Shaposhnikov:2008xi,Gretsch:2013ooa,Garcia-Bellido:2011kqb,Bezrukov:2012hx}. The reason of the presence of quantum scale anomalies is related to the absence of regularization, which preserves scale invariance. If we use the standard dimensional regularization, the interacting Lagrangian is not scale-invariant when the spacetime dimension deviates from~4. However, it is possible to define a theory in $d$ dimensions in such a way that the scale symmetry is kept explicit. For example, the Standard Model can be extended to the Higgs-dilaton model with spontaneously broken scale invariance by the price of adding an extra scalar field \cite{Shaposhnikov:2008xb}
\begin{equation}
\label{HD}
\frac{\mathcal{L}}{\sqrt{-g}}=-\frac{1}{2}\left(\xi_{\chi}\chi^2+2\xi_h H^{\dagger} H\right)R+\frac{1}{2}(\partial \chi)^2+ \partial H^{\dagger}\partial H-\frac{\lambda}{4}\chi^{\frac{4-d}{d-2}}(H^{\dagger}H-\alpha^2 \chi^2)^2+\mathcal{L}_{SM}. 
\end{equation}
The field $\chi$, which is called dilaton, receives the vacuum expectation value (VEV) \mbox{$\langle\chi \rangle=v/\alpha$}, and the rest of the SM Lagrangian $\mathcal{L}_{SM}$ gets modified similarly by multiplying by the appropriate power of the dilaton field. If $\xi_{\chi} \langle\chi \rangle^2=M_P^2$\footnote{We assume that the dilaton VEV is much larger than that of the Higgs, i.e. $\alpha\ll 1$. }, then the Planck scale would arise from the dilaton VEV. At the level of quantum theory, the renormali\-zation scale, introduced in the original Standard Model, is replaced by the dilaton VEV. This cancels the scale anomaly preserving the quantum scale symmetry\footnote{In flat space, even the broader symmetry group -- the conformal symmetry -- can remain anomaly-free. However, in the presence of dynamical gravity, this symmetry reduces to the scale invariance~\cite{Shaposhnikov:2022dou,inpreparation}.}. A serious drawback of such theories is that, in opposite to the original Standard Model, they are non-renormalizable even without gravity and can be treated only as effective field theories that work at energies $E \ll \langle \chi \rangle$ \cite{Shaposhnikov:2009nk}. At higher energies, they should match some strongly coupled scale invariant theory \cite{Karananas:2017zrg}. If all the phenomena we aim to describe happen at the scales much smaller than the dilaton VEV, the framework of spontaneously broken scale symmetry is still predictive. 

One of the distinctive features of the low energy theories arising from spontaneous breaking of scale symmetry is the presence of the massless dilaton --- the Goldstone boson of the broken symmetry. The metric field redefinition $g_{\mu\nu}\rightarrow \Omega(\chi, h)^2 \hat{g}_{\mu\nu}$\footnote{Hereafter we use the unitary gauge, where $H=(0,h/\sqrt{2})^T$.} brings the Lagrangian to the so-called Einstein frame, where gravity is described by the Einstein-Hilbert action. Further canonical normalization of the scalar fields leads to the action of the general form
\begin{equation}
\label{HD_gen}
    S=\int d^4 x \sqrt{-\hat{g}}\left(-\frac{1}{2}M_P^2 \hat{R}+\frac{1}{2}F(\phi)(\partial r)^2+\frac{1}{2}(\partial \phi)^2-V(\phi)\right).
\end{equation}
Here the canonically normalized field $\phi$ is non-linearly related to the Higgs, $F(\phi),~V(\phi)$ are functions of $\phi$, and $r$ is a dimensionless shift-symmetric Goldstone field of the broken scale invariance (we will also call it dilaton hereafter). The scale symmetry of the original action, therefore, translates into the shift symmetry of the dilaton field in the Einstein frame. The detailed procedure of the field redefinitions bringing \eqref{HD} to the form \eqref{HD_gen} is described, for example in \cite{Garcia-Bellido:2011kqb,Garcia-Bellido:2012npk,Bezrukov:2012hx}, where this model was studied in the cosmological context of Higgs-driven inflation.

Regardless of the concrete model, the action \eqref{HD_gen} describes a low-energy theory of two scalar fields, where one of them is a Goldstone of spontaneously broken scale symmetry, which was present in the corresponding Jordan-frame formulation. This expression demonstrates all possible terms with no more that 2 derivatives that can be written for two scalars. What are the other couplings between the dilaton and the SM particles? It can have only derivative couplings to the matter fields, which makes the fifth force constraints~\cite{Wetterich:1987fm, Wetterich:1987fk,Shaposhnikov:2008xb,Ferreira:2016kxi} satisfied. 

There is an interesting possibility claimed in \cite{Shaposhnikov:2022dou,inpreparation} that if we make the field $r$ transforming as a pseudoscalar with respect to parity transformations, then this field can be coupled to the topological density of the SM gauge fields, which can lead to non-perturbative breaking of the scale symmetry. In particular, the presence of this coupling to gluons would modify the QCD Lagrangian as
\begin{equation}
\label{LQCDr}
\mathcal{L}_{QCD}=-\frac{1}{16\pi\alpha_s}G_a^{\mu\nu}G^a_{\mu\nu} +\gamma r \, {\tilde G}_a^{\mu\nu}G^a_{\mu\nu}. 
\end{equation}
Here $\gamma$ is a dimensionless constant. The key observation relevant for this work is that after the canonical normalization of the dilaton field $r$, it enters the Lagrangian precisely in the same way as the axion field introduced for resolving the strong CP-problem, see~\cite{Marsh:2015xka,GrillidiCortona:2015jxo} for a comprehensive review. 

The idea of identifying the axion with the dilaton opens the possibility to connect the scale invariance in the Jordan frame and the axion solution to the strong CP-problem in QCD, without actual introducing a special degree of freedom. The dilaton interaction with the strongly coupled QCD at low energies leads to non-perturbative breaking of the shift symmetry and to the appearance of the potential identical to that in the standard QCD axion scenario. In this case, the dilaton with the non-perturbatively generated mass can form dark matter. The main purpose of this work is to explore the possibility for the dilaton, coupled to the QCD topological term, to form dark matter in late Universe. We construct an explicit model, discuss the conditions under which it is consistent with the early Universe inflation driven by the Higgs field, and find the allowed parameter region, where the dilaton can be dark matter in the late Universe.\\

The paper is organised as follows. In Section~\ref{sec:II}, we discuss the idea of employing the dilaton field to solve the strong CP-problem in QCD and explain dark matter at the same time. In Section~\ref{sec:III}, we introduce the model of Higgs-driven inflation, which is capable to generate the dilaton dark matter via the same mechanism as in the QCD axion models. In Section~\ref{sec:IV}, we obtain the CMB constraints on inflation and bound the parameters of the model presented in Section~\ref{sec:III}. In Section~\ref{sec:V}, we discuss additional isocurvature constraints and show the parameter space, for which the dilaton dark matter is compatible with inflation. In Section~\ref{sec:VI}, we summarise the results.


\section{Can dilaton solve strong CP-problem and form dark \mbox{matter?}}\label{sec:II}

The QCD sector of the SM Lagrangian should also contain non-zero CP violating term
\begin{equation}
\label{L}
    \mathcal{L}_{CP} = \frac{\alpha_s}{8\pi}\,\theta_0{\tilde G}_a^{\mu\nu}G^a_{\mu\nu}
\end{equation}
This term gives an impact on measurable quantities only in a combination with the phase of the determinant of the CKM quark matrix. The bound on the neutron dipole moment implies \cite{Schmidt-Wellenburg:2016nfv,Pendlebury:2015lrz}
\begin{equation}
\label{theta}
    \Delta=\theta_0+ {\rm Arg~ det~} M_q <10^{-10}.
\end{equation}
Since the elements of the CKM matrix $M_q$ are of order unity, the cancellation at the level $10^{-10}$ in (\ref{theta}) looks surprising. This well-known fine-tuning problem for the value of $\theta_0$ arising in the Standard Model is usually called the strong CP-problem. 

The commonly discussed solution to the strong CP-problem first proposed in~\cite{Peccei:1977hh} involves adding an extra pseudoscalar particle called the axion. The low-energy effective Lagrangian for this particle reads~\cite{Peccei:1977hh}
\begin{equation}
\label{L_a}
L=\frac{1}{2}(\partial a)^2 +\frac{a}{f_a}\frac{\alpha_s}{8\pi}{\tilde G}^{\mu\nu}G_{\mu\nu}+\frac{g_{a\gamma\gamma}\, a}{f_a}{\tilde F}^{\mu\nu}F_{\mu\nu}+\frac{\partial_{\mu}a}{2 f_a}J_{a}^{\mu}
\end{equation}
Here $F_{\mu\nu}$ is the photon tensor, $g_{a\gamma\gamma}$ is the axion-photon coupling, and $J_{a}^{\mu}$ is a model-dependent combination of the SM quark axial currents. The axion $a$ has a shift symmetry, which allows to absorbe a non-zero value of the $\theta_0$-angle into the redefinition of the field. 

The term $a\,{\tilde G}^{\mu\nu} G_{\mu\nu}$ has a form of the (non-perturbative) effective action provided by the chiral anomaly. Thus, it can be absorbed by the quark sector within the appropriate chiral rotation. In the strong coupling regime at zero temperature, the QCD sector is well described by the chiral Lagrangian for pions. The latter chiral rotation corresponds to the well-defined transformation of pion fields in the chiral Lagrangian. Thus, the axion will be mixed with pions leading to the appearance of the axion potential \cite{DiVecchia:1980yfw}
\begin{equation}
\label{V}
V(a)=m_{\pi}^2 f_{\pi}^2\sqrt{1-\frac{4 m_u m_d}{(m_u+m_d)^2}\sin^2{\left(\frac{a}{2 f_a}\right)}},
\end{equation}
which corresponds to the axion mass
\begin{equation}
m_a^2=\frac{ m_u m_d}{(m_u+m_d)^2}\frac{m_{\pi}^2 f_{\pi}^2}{f_a^2}.
\end{equation}
Here $f_{\pi}$ is a pion decay constant, and $m_u$, $m_d$ are the $u$ and $d$ quark masses, respectively. The appearance of the axion mass in non-perturbative regime leads to vanishing of $\Delta$ dynamically. However, the axion shift symmetry gets broken explicitly, due to non-perturbative effects. The discrete symmetry $a\rightarrow a+2\pi f_a$ remains unbroken, as it can be seen from~\eqref{V}.

The scale-invariant Lagrangian \eqref{HD_gen} formulated in the Einstein frame, together with the QCD terms~\eqref{LQCDr}, can be brought to the form identical to (\ref{L_a}) by means of canonical normalization of the field $r$ in the SM vacuum. Therefore, the dilaton Lagrangian also provides a solution to the strong CP-problem, if the dilaton is defined as a pseudoscalar in the Einstein frame. Namely, it transforms under the parity transformation as $r \rightarrow -r$, and is coupled to the term $\tilde{G}^{\mu\nu}G_{\mu\nu}$, thus preserving the CP symmetry. Such pseudoscalar dilaton would correspond to some generalized scale symmetry supplemented by the parity transformation.

In the non-perturbative regime of QCD, the dilaton potential is generated in the same manner as it happens in the axion models. It breaks the continuous scale invariance to a discrete subgroup. This breaking again guarantees zero value of $\Delta$ in the vacuum, thus solving the strong CP-problem. Moreover, all the low energy phenomenology and cosmology of the dilaton would be identically the same as in the axion models, where everything is determined by the scale $f_a$. 
The standard cosmological axion dark matter scenarios are based on the spontaneous breaking of the $U(1)$ Peccei-Quinn (PQ) symmetry \cite{Peccei:1977hh}. The low energy Lagrangian \eqref{L_a} comes from a renormalizable action, where the axion corresponds to the phase of the Peccei-Quinn field. The axion realization in a renormalizable model also requires the presence of the extra quarks (KSVZ model) \cite{Kim:1979if,Shifman:1979if} or the extra Higgs doublet (DFSZ model) \cite{Dine:1981rt}. The distinctive feature of the minimal scenario under consideration is that we do not build a UV completion of the axion Lagrangian, embedding it to a renormalizable model. Instead of that, we use an effective axion Lagrangian and expect that at the energy scale $f_a$ the model flows to some strongly coupled  theory, where the exact quantum scale invariance is restored. Although we do not make an explicit construction for that type of UV completion, we assume that all phenomena under consideration, including inflation, take place at energies below $f_a$. If this condition is satisfied, the model becomes predictive.
The most common cosmological scenario for the axion dark matter is constructed under the assumption that the PQ symmetry, being broken during inflation, was thermally restored at the reheating stage~\cite{Dine:1981rt,Abbott:1982af,Preskill:1982cy}. During cooling down of the SM plasma, the PQ symmetry gets broken again. Below the QCD phase transition temperatures, the axion acquires a potential. Given that the initial value of the axion field $a$ was distributed randomly between $0$ and $2\pi f_a$, the axion initial displacement from the minimum of the potential is likely to be of order $f_a$ (i.e. $\theta_0 \sim 1$). Thus, the axion field starts to oscillate and perfectly mimics the dark matter component in late Universe. 

In our model, scale invariance plays the role of the PQ symmetry, which means that the theory is well-defined only in the broken phase. For this reason, we can consider only the scenario, in which PQ symmetry is \textit{always} broken. Such models are typically not compatible with high-scale inflation, because the axion generates large isocurvature perturbations \cite{Lyth:1989pb}. The latter are tightly constrained by the CMB data~\cite{Planck:2018jri,Planck:2018vyg}, which motivates to search for models with low-scale inflation~\cite{Takahashi:2018tdu}. In the model where the dilaton plays the role of the axion, there is a natural mechanism of suppression the isocurvature modes. 

Isocurvature bounds on the axion models are derived under the assumption that the axion is \textit{not} coupled to the inflaton field. Clearly, in the Lagrangian~\eqref{HD_gen}, the normalization of the axion field depends on the value of the inflaton, which can significantly differ at inflation and in the SM vacuum. If the effective axion constant $f_{a,I}$ during inflation has a value, which significantly exceeds $f_{a,0}$, the isocurvature fluctuations get suppressed by the factor $f_{a,0}/f_{a,I}\ll 1$. In numbers, the requirement that all dark matter in the late Universe be made up of axions almost fixes the value of $f_{a,0}$
\begin{equation}
    f_{a,0}=2.3\cdot 10^{-7} \,M_P \,\langle\theta_\text{init}^2\rangle^{-6/7}.
\end{equation}
Here $\theta_\text{init}^2$ is the initial axion displacement in terms of the QCD $\theta$-angle. To avoid fine-tuning, we assume $\theta_\text{init}^2\sim 1$, thus fixing this scale to be a bit smaller than the Hubble scale in the large-field inflation models. For this reason, we need to arrange inflation where the Hubble scale $H_I$ is lower than $f_{a,0}$. Since the Hubble scale is defined by the CMB amplitude $A_s=2.2\cdot 10^{-9}$ and the tensor-to-scalar ratio $r_t$
\begin{equation}\label{eq:low-Hubble}
    H_I^2 = \frac{V_E(\phi_I)}{3 M_P^2} = \frac{1}{2}\pi^2 r_t A_s M_P^2,
\end{equation}
the condition $H_I<f_{a,0}$ translates to the upper bound on $r_t$
\begin{equation}\label{rt-Introduction}
    r_t\lesssim 5 \cdot 10^{-6},
\end{equation}
which is much tighter than the Planck bound $r_t<0.05$~\cite{Planck:2018jri,Planck:2018vyg}. 
Following the results of \cite{Kearney:2016vqw}, suppression of the axion isocurvature perturbations requires\footnote{Recent simulations \cite{Ballesteros:2021bee} show that the isocurvature perturbations can grow during reheating making this bound model-dependent. We leave the question whether the results of our model can get affected by the similar effects for future study.}
\begin{equation}\label{iso_cond}
    \frac{f_{a,I}}{f_{a,0}} > 1.6 \cdot 10^5\sqrt{\frac{r_t}{5\cdot 10^{-6}}}.
\end{equation}
To summarise, let us list the conditions, which should be satisfied for a viable model of the dilaton dark matter~\eqref{HD_gen} generated in the same way as in the axion models. Here we assume that $\phi$ in~\eqref{HD_gen} stands for the inflaton field\footnote{Although we consider Higgs-driven inflation, the model can also work with other mechanisms of inflation.}.
\begin{itemize}
    \item All relevant energy scales (such as Hubble scale of inflation) should be smaller than the axion constant $f_{a,0}$, which is fixed by the dark matter abundance. This condition prefers low scale inflation.
    \item The potential $V(\phi)$ of the inflaton should provide the stage of slow-roll inflation leading to the parameters of the perturbation spectrum $n_s,~r_t$ compatible with the Planck data.
    \item Isocurvature perturbations are to be suppressed by the requirement~\eqref{iso_cond}. Recall that $f_{a,I}/f_{a,0}=\sqrt{F(\phi_I)/F(\phi_0)}$ (see~\eqref{HD_gen}), where $\phi_0$ is the value of the inflaton field in the late-time low-energy vacuum.
\end{itemize}
Under these conditions, the proposed scenario of identifying the dilaton and the QCD axion would be a viable model describing inflation and dark matter formation in the late Universe. In the upcoming sections, we construct an explicit example of such a model formulated in the Jordan frame, which meets all these conditions.


\section{Dilaton dark matter in Higgs-driven inflation}\label{sec:III}

In this section, we present the Jordan-frame formulation of the scale-invariant model realising Higgs-driven inflation with the dilaton dark matter. We start from the following Jordan-frame action
\begin{equation}\label{eq:SJ}
    S_J = \int d^4 x\sqrt{|g|}\,\Big\{\mathcal{L}_\text{grav} + K_\text{HD} - V_\text{HD} + \mathcal{L}_\text{QCD}\Big\},
\end{equation}
where
\begin{equation}\label{eq:Lgrav}
    \mathcal{L}_\text{grav} = -\frac{R}{2}\left(\xi_\chi \chi^2 + \xi_h h^2 + \xi\frac{h^4}{\chi^2}\right),
\end{equation}
is the gravity sector of the theory, 
\begin{equation}\label{eq:KHD}
    K_\text{HD} = \frac{1}{2}(\partial\chi)^2\left(1 + \alpha\frac{h^2}{\chi^2}\right) + \frac{1}{2}(\partial h)^2\left(1 + \beta\frac{h^2}{\chi^2}\right),
\end{equation}
is the Higgs-dilaton kinetic part,
\begin{equation}\label{eq:VHD}
    V_\text{HD} = \frac{\lambda}{4}\left(h^2 - \delta_0\chi^2\right)^2\left(1 + \delta_1\frac{h^2}{\chi^2} + \delta_2\frac{h^4}{\chi^4}\right),
\end{equation}
is the Higgs-dilaton potential, and
\begin{equation}\label{eq:LQCD}
    \mathcal{L}_\text{QCD} = -\frac{1}{16\pi \alpha_s}G^{\mu\nu}G_{\mu\nu} + \frac{\gamma}{64\pi^2}\log\left[\frac{\chi^2 + h^2}{M_P^2}\right]\tilde{G}^{\mu\nu}G_{\mu\nu},
\end{equation}
is the QCD Lagrangian including the topological term coupled to the Higgs and the dilaton. New introduced parameters are as follows:
\begin{itemize}

    \item non-minimal couplings $\xi_\chi$, $\xi_h$ and $\xi$ of the dilaton, the Higgs and their mixed term with gravity, respectively;
    
    \item the coefficients $\alpha$ and $\beta$ of the scale-invariant corrections to the kinetic terms of the fields;
    
    \item the vacuum constant $\delta_0$, which along with the expectation value of $\chi$ defines the Higgs vacuum expectation value~$v$;
    
    \item the corrections to the Higgs-dilaton potential $\delta_1$ and $\delta_2$, which determine the spectral characteristics of inflation;
    
    \item the coupling $\gamma$ to the topological QCD-term.
    
\end{itemize}

We assume $\langle\chi\rangle\sim M_P$, so that the main contribution to the Planck mass $M_P$ in the gravity term~\eqref{eq:Lgrav} comes from the dilaton $\chi$, and ordinary GR is recovered in the low-energy limit. Also, under this assumption, all the corrections in~\eqref{eq:KHD} and~\eqref{eq:VHD} are negligible around the Higgs vacuum $\langle h \rangle \sim v$, i.e. $\displaystyle\frac{h^2}{\chi^2} \sim \frac{v^2}{M_P^2} \ll 1$. They might be thought of as the leading-order UV-corrections to the EFT of the Higgs-dilaton inflation. All the corrections are quadratic in fields, since we are to respect the $\mathbb{Z}_2$~symmetry~$h \rightarrow -h$, inherited in the unitary gauge from the $SU_L(2) \times U_Y(1)$ symmetry of the Higgs field. The powers of the dilaton serve to make the whole action scale invariant.\\

Hereafter we choose to work in the Palatini formalism, in which the metric tensor and the connection are treated as independent fields. Even though both the metric formalism and the Palatini give the same vacuum Einstein equations, they differ in the case when matter fields depend on the connection, e.g. in scalar-tensor theories. Higgs and Higgs-dilaton inflations are examples of such models ~\cite{Almeida:2018oid,Rubio:2019ypq,Shaposhnikov:2020fdv,Rasanen:2022ijc,Ito:2021ssc,Rasanen:2018ihz,Enckell:2020lvn,Piani:2022gon,Yin:2022fgo}. 

The choice of the Palatini formulation is not arbitrary. In the standard metric scenario, the Higgs field is strongly coupled to the Ricci scalar, and the consequence of this interaction is that the UV cutoff of the Higgs sector is of the same order of magnitude as the Hubble rate during inflation~\cite{Bezrukov:2010jz}. It makes the theory dependent on the unknown ultraviolet completion of the Higgs sector. On the contrary, in Palatini gravity, Higgs inflation does not suffer from the unitarity violation~\cite{Shaposhnikov:2020fdv}, since the UV-cutoff is much higher than the inflation energy scale, and the unknown high-energy physics does not contribute to the inflationary dynamics.\\

Performing the Weyl trasformation of~\eqref{eq:SJ} with the factor $e^{2\varrho}$
\begin{equation}\label{eq:Weyl}
    g_{\mu\nu}\rightarrow e^{2\varrho} g_{\mu\nu}, \qquad e^{2\varrho} = \frac{\xi_h h^2\chi^2 + \xi h^4 + \xi_\chi\chi^4}{\chi^2 M_P^2},
\end{equation}
and introducing new field variables as
\bea
    \phi(\varphi) &=& \frac{M_P}{2}\int\limits^{\varphi}_0 ds\,\frac{|\sec s|^3\Big(\alpha + 1 - (\alpha - 1)\cos2s\Big)}{\sqrt{\Big(\alpha\tan^2s + 1\Big)\Big(\xi\tan^4s + \xi_h\tan^2s + \xi_\chi\Big)}}, \label{eq:NewFieldF} \\
    r(\rho) &=& \log\frac{\rho}{M_P} \label{eq:NewFieldr},
\eea
where
\begin{equation}
    \rho^2 = h^2 + \chi^2, \qquad \varphi = \arctan\frac{\chi}{h},
\end{equation}
we obtain the Einstein-frame action
\begin{equation}\label{eq:Ldiag}
    \begin{aligned}
         &S_E = \int d^4x\sqrt{|\hat{g}|}\left[-\frac{\hat{R}}{2}M_P^2 + \frac{1}{2}M_P^2\,\frac{\sec^2\varphi\left(\alpha\sec^2\varphi - \alpha + 1\right)}{\xi\tan^4\varphi + \xi_h\tan^2\varphi + \xi_\chi}\left(\partial r\right)^2 + \frac{1}{2}\left(\partial\phi\right)^2\right.\\
         &\left. - \frac{\lambda M_P^4}{4}\,\frac{\sin^4\varphi(\phi)\Big(\delta_1\tan^2\varphi(\phi) + \delta_2\tan^4\varphi(\phi) + 1\Big)}{\Big(\sin^2\varphi(\phi)\left(\xi\tan^2\varphi(\phi) + \xi_h\right) + \xi_\chi\cos^2\varphi(\phi)\Big)^2} \right.\\
         &\left. - \frac{1}{4g^2}G^{\mu\nu}G_{\mu\nu} - \frac{\gamma}{32\pi^2}\,r\,\tilde{G}^{\mu\nu}G_{\mu\nu}\right],
    \end{aligned}
\end{equation}
where $\varphi(\phi)$ is given by the inverse of~\eqref{eq:NewFieldF}. We have taken $\alpha = \beta$ so that only the field~$r$ is coupled to the topological QCD term. This choice makes this model similar to the one that describes the QCD axion~\cite{Peccei:1977hh,Marsh:2015xka} (see Appendix~\ref{sec:appendixA} for details). In this construction, the dilaton naturally becomes the axion suitable for solving the strong CP-problem.

In the Einstein frame, the scalar sector is minimally coupled to gravity at the cost that the kinetic term of the dilaton $r$ is no longer canonically normalized. The Higgs is now represented by the field~$\phi$ which plays the role of the inflaton --- see Appendix~\ref{sec:appendixA} for a detailed discussion.

In terms of the new field variables~\eqref{eq:NewFieldF} and~\eqref{eq:NewFieldr}, the initial Higgs-dilaton potential~\eqref{eq:VHD} does not depend on $r$ (see~\eqref{eq:Ldiag}). This means that $r$ is a massless Goldstone mode of the corresponding scale invariance, which was explicitly broken by the Planck mass $M_P$ after performing the Weyl transformation~\eqref{eq:Weyl}.


\section{Palatini Higgs-dilaton inflation}\label{sec:IV}

Our general assumption is that the vacuum expectation value of the dilaton $\chi$ is of order of the Planck scale $\langle\chi\rangle\simeq M_P$, so when the Higgs takes super-Planckian values $h \gg M_P$, we have
\begin{equation}\label{eq:super-Planck}
    h \gg \chi \qquad \text{at inflation.}
\end{equation}
Let us consider the Einstein-frame potential in terms of the initial fields $h$ and $\chi$ (see~\eqref{einstein_lagrangian} in Appendix~\ref{sec:appendixA})
\begin{equation}\label{eq:VHDE}
    V_E = \frac{\lambda M_P^4}{4}\,\frac{\displaystyle h^4\left(1 + \delta_1\frac{h^2}{\chi^2} + \delta_2\frac{h^4}{\chi^4}\right)}{\displaystyle\left(\xi_h h^2 + \xi\frac{h^4}{\chi^2} + \xi_\chi \chi^2\right)^2}.
\end{equation}
At super-Planckian scales~\eqref{eq:super-Planck}, the leading-order contribution in the numerator is that determined by the coupling $\delta_2$
\begin{equation}\label{eq:VE-dominant-term}
    V_E \simeq \frac{\lambda M_P^4}{4}\,\frac{\delta_2 h^4}{\left(\xi_h \chi^2 + \xi h^2 + \xi_\chi\displaystyle\frac{\chi^4}{h^2}\right)^2}.
\end{equation}
We also see that the last term in the denominator is suppressed at inflation. Moreover, in our case, the inflationary dynamics is determined not by the Higgs-gravity coupling~$\xi_h$, as in the standard Higgs or Higgs-dilaton inflation scenarios, but by the mixed-field one $\xi$, because the potential flattens at the value
\begin{equation}
    V_E\left(h \gg \chi\right) \simeq \frac{\lambda\delta_2 M_P^4}{4\xi^2}.
\end{equation}
This opens the window to interesting inflationary dynamics, into which the dilaton is implicitly embedded through the coupling~$\xi$. 

\begin{figure}[h!]
\center{\includegraphics[width=0.55\linewidth]{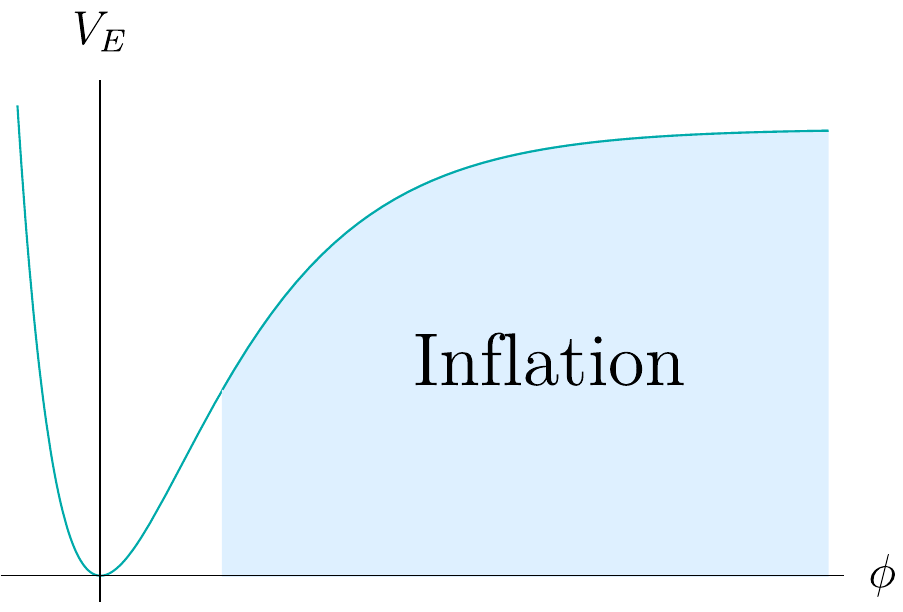}}
\caption{Inflationary potential~\eqref{eq:in-the-main-text-pot} as a function of the canonical Higgs-inflaton $\phi$. The region of the potential where inflation happens is shown in blue (see Appendix~\ref{sec:appendixB}).}
\label{fig:inflation-region}
\end{figure}

Since we are especially interested in a scenario, in which the field $\chi$ plays a subdominant role during inflation: $\chi \ll h$, one can see from~\eqref{eq:VE-dominant-term} that we should require the following hierarchy
\begin{equation}\label{eq:hierarchy}
    \xi_h \ll \xi.
\end{equation}

Under the condition~\eqref{eq:hierarchy}, the inflationary Starobinsky-like effective potential can be obtained from~\eqref{eq:VE-dominant-term} in the following form (see Appendix~\ref{sec:appendixB} for derivation)
\begin{equation}\label{eq:in-the-main-text-pot}
    V_E(\phi) \simeq \frac{\lambda\delta_2 M_P^4}{4\xi^2}\left(1 - \frac{\xi_h}{\xi}\sinh^2\left[\sqrt{\frac{\xi}{\alpha}}\frac{\phi - \phi_\text{vac}}{M_P}\right]\right)^{2},
\end{equation}
where $\phi$ is the canonical inflaton. This potential with the region where inflation happens is shown in Fig.~\ref{fig:inflation-region}. 

The flattening of the potential~\eqref{eq:in-the-main-text-pot} at super-Planckian scales of $\phi$ allows for the slow-roll regime. The slow-roll parameters are given by
\begin{equation}\label{eq:eps-gen-main-text}
    \begin{aligned}
        &\epsilon \equiv \frac{M_P^2}{2}\left(\frac{V_E^\prime(\phi_I)}{V_E(\phi_I)}\right)^2 \simeq \frac{4\xi_h^2}{-\alpha\xi + 2\xi_h^2\sqrt{\displaystyle\frac{2\alpha\xi}{\xi_h^2} + 4}\sinh\left(\displaystyle\frac{8\xi_h N_e}{\alpha}\right) + \left(\alpha\xi + 4\xi_h^2\right)\cosh\left(\displaystyle\frac{8\xi_h N_e}{\alpha}\right)},\\
        &\eta \equiv M_P^2\frac{V_E^{\prime\prime}(\phi_I)}{V_E(\phi_I)} \simeq \frac{8\xi_h^3\sqrt{\displaystyle\frac{2\alpha\xi}{\xi_h^2} + 4} - 4\alpha\xi\xi_h\sinh\left(\displaystyle\frac{8\xi_h N_e}{\alpha}\right)}{\alpha\left(-\alpha\xi - 4\xi_h^2 + \alpha\xi\cosh\left(\displaystyle\frac{8\xi_h N_e}{\alpha}\right)\right)}.
    \end{aligned}
\end{equation}

The statistical information is predominantly encoded in the two-point functions of scalar and tensor modes of primordial perturbations -- the power spectra. The spectra are parameterized to probe the deviations from the scale invariant form as follows
\begin{equation}
    P_s = A_s\left(\frac{k}{k_0}\right)^{n_s - 1}, \qquad P_t = A_t\left(\frac{k}{k_0}\right)^{n_t - 1}, \qquad r_t \equiv \frac{A_t}{A_s} = 16\epsilon.
\end{equation}
Here $k_0=0.002$ Mpc is the reference scale for the Planck observations~\cite{Planck:2018vyg}. The spectral index $n_s$ is given by
\begin{equation}
    n_s = 1 + 2\eta - 6\epsilon.
\end{equation}
The analytical expression for the amplitude of scalar perturbation has the form
\begin{equation}\label{eq:As-main-text}
    A_s = \frac{V_E(\phi_I)}{24\pi^2\epsilon M_P^4} \simeq \frac{\lambda\delta_2}{1536\pi^2\xi^4\alpha N_e}(\alpha - 8\xi N_e)^2\left(2\xi N_e + \sqrt{2\alpha\xi}\right),
\end{equation}
where $N_e \sim 60$ is the number of e-folds.

The amplitude of scalar perturbations $A_s$ fixes the height of the potential $\delta_2$. Given the observational constraint on the normalization of the primordial spectrum at large scales~\cite{Planck:2018vyg}
\begin{equation}\label{eq:As-Planck}
    \log\left(10^{10}A_s\right) \simeq 3.094 \pm 0.034,
\end{equation}
we have from~\eqref{eq:As-main-text} the approximate expression for $\delta_2$ in terms of $\alpha$ and $\xi$
\begin{equation}\label{eq:delta-2}
    \delta_2 \simeq \frac{0.1\,\alpha\,\xi^4}{\left(120\xi + \sqrt{\alpha\xi}\right)\left(\alpha - 480\,\xi\right)^2}.
\end{equation}

The spectral tilt is bounded at the level~\cite{Planck:2018vyg}
\begin{equation}\label{eq:ns-Planck}
    n_s = 0.9661 \pm 0.0040,
\end{equation}
while the tensor-to-scalar ratio~\cite{Planck:2018vyg}
\begin{equation}\label{eq:rt-Planck}
    r_t < 0.05.
\end{equation}

\begin{figure}[h!]
\center{\includegraphics[width=0.55\linewidth]{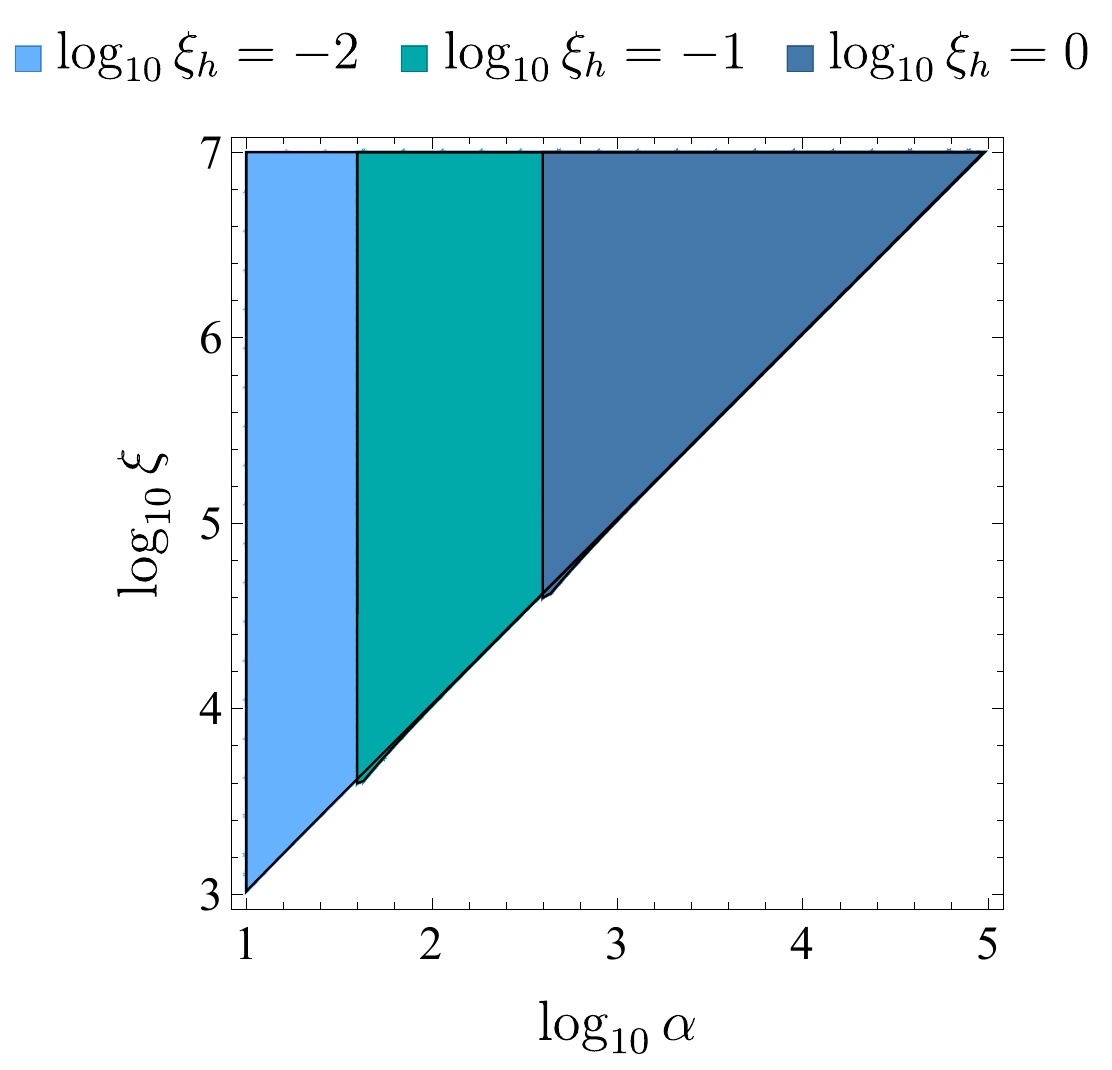}}
\caption{Region plot of the allowed parameter space $(\xi, \xi_h, \alpha)$ in logarithmic scale as a result of the Planck constraints on $n_s$~\eqref{eq:ns-Planck} and $r_t$~\eqref{eq:rt-Planck}. Here, the requirement of non-violation of unitarity~\eqref{eq:fa0-cutoff} is also taken into account. The hierarchy~\eqref{eq:hierarchy} is assumed. The values of $\xi_h$ are fixed and shown on the top. We see that as $\xi_h$ increases, larger values of $\xi$ and $\alpha$ are needed to meet the CMB constraints.}
\label{fig:inflation-constraints}
\end{figure}

Using~\eqref{eq:eps-gen-main-text} for the calculation of $n_s$ and $r_t$ and combining them with the observational constraints~\eqref{eq:ns-Planck} and~\eqref{eq:rt-Planck}, we obtain the allowed parameter space, which is shown in Fig.~\ref{fig:inflation-constraints}. 

\section{Isocurvature constraints on dilaton dark matter}\label{sec:V}

Axion models assuming Peccei-Quinn symmetry breaking during inflation are in tension with the high scale inflationary models, because it is hard to suppress the axion isocurvature mode. In the model under consideration, the role of the Peccei-Quinn symmetry is played by the scale symmetry, which cannot be described by the proposed effective theory in the unbroken phase. Therefore, this symmetry is \textit{always} broken. However, the parameter space (see Fig.~\ref{fig:inflation-constraints}) allows for a low Hubble scale during inflation~\eqref{eq:low-Hubble} with tiny $r_t$, which gives us hope to avoid the axion isocurvature tension and fine-tuning issues.\\

Let us start with the preliminaries related to the axion dark matter. Axion gains a potential due to non-perturbative QCD effects, which depend on the temperature of the SM plasma. Being massive in the late universe, the oscillating axion condensate can be a constituent part of the dark matter. Its abundance $R_a$ is given by~\cite{Visinelli:2014twa}
\begin{equation}
    R_a\equiv\frac{\Omega_a}{\Omega_{CDM}} = 5.6 \cdot 10^7 \langle r^2 \rangle \left(\frac{f_{a,0}}{M_P}\right)^{7/6}.
    \label{eq:DM}
\end{equation}
Here we denote the axion normalization constant in the present Higgs vacuum $\phi_{\text{vac}} = 0$ as $f_{a,0}$; $\langle r^2 \rangle = r_\text{init}^2 + \delta r^2$ is the total root mean square displacement of the axion from the QCD minimum (in terms of the angular variable $r = a/f_a$, where $a$ is the canonical axion field), which might contain the initial displacement $r_\text{init}$ and the fluctuation~$\delta r$ (neglecting anharmonic factors~\cite{Turner:1985si}). Note that the relation~\eqref{eq:DM} fixes the value of $f_{a,0}$ as
\begin{equation}
    \frac{f_{a,0}}{M_P} \simeq 2.3 \cdot 10^{-7}, 
\end{equation}
if the axion constitutes all observable dark matter: $R_a \sim 1$, and its displacement takes a natural value: $\langle r^2 \rangle \sim 1.$ Roughly, the value~\eqref{eq:fa0-cutoff} defines the UV cutoff of the theory, so we want the Hubble scale at inflation~\eqref{eq:low-Hubble} to be lower than $f_{a,0}$
\begin{equation}\label{eq:fa0-cutoff}
    H_I \lesssim f_{a,0}.
\end{equation}
We use this unitarity constraint in addition to the isocurvature bound. Taking into account the measured value of the amplitude of scalar perturbations $A_s$~\eqref{eq:As-Planck}, this relation implies that the tensor-to-scalar ratio $r_t$ should be additionally suppressed at the level (see~\eqref{eq:low-Hubble} and~\eqref{rt-Introduction})
\begin{equation}
    r_t \lesssim 5 \cdot 10^{-6}.
\end{equation}
In Fig.~\ref{fig:inflation-constraints}, this bound is taken into account.

As we have mentioned, the scale invariance is always broken, therefore, at inflation the axion-dilaton is already massless, and the symmetry is not restored by quantum fluctuations of the inflaton or by thermal fluctuations during reheating. Under these conditions, such axion produces uncorrelated with adiabatic modes isocurvature fluctuations~\cite{Marsh:2015xka}. Due to the non-canonical kinetic term of the axion-dilaton, the value of the normali\-zation constant $f_{a,I}$ at inflation differs from that in the SM vacuum. The isocurvature fluctuation $\delta a$ reads~\cite{Linde:1991km,Linde:1990yj,Lyth:1989pb}
\begin{equation}\label{eq:isocurvature}
    \delta a = \frac{H_I}{2\pi}\frac{f_{a,0}}{f_{a,I}}.
\end{equation}

The axion constant is given by (see~\eqref{3} in Appendix~\ref{sec:appendixA})
\begin{equation}
    f_a(\varphi) = \frac{M_P}{\gamma}\,\sqrt{\frac{\sin^2\varphi\left(\alpha+\beta\tan^2\varphi+1\right)+\cos^2\varphi}{\sin^2\varphi\left(\xi\tan^2\varphi+\xi_h\right)+\xi_\chi\cos^2\varphi}}.
\end{equation}

At inflation and in the case $\alpha = \beta$, $f_{a,I} \equiv f_a(\varphi_I)$ reduces to (see~\eqref{inflation} in Appendix~\ref{sec:appendixB})
\begin{equation}\label{faI}
    f_{a,I} = \frac{M_P\sqrt{\alpha}}{\gamma\sqrt{\xi\sin^2\varphi_I + \xi_h\cos^2\varphi_I}},
\end{equation}
while in the SM vacuum $\varphi = 0$ it is given by
\begin{equation}\label{eq:fa0}
    f_{a,0} \equiv f_a(0) = \frac{M_P}{\gamma\sqrt{\xi_\chi}}.
\end{equation}

Taking as an estimate $\varphi_I \sim \displaystyle\frac{\pi}{2}$, we obtain the following expression for the ratio of~\eqref{faI} and~\eqref{eq:fa0}
\begin{equation}\label{eq:fa-ratio}
    \frac{f_{a,I}}{f_{a,0}} \simeq \sqrt{\frac{\alpha\xi_\chi}{\xi}}.
\end{equation}

Careful estimation of~\eqref{eq:fa-ratio} takes into account the more accurate value of the angular variable $\phi_I$ on inflation, which can be found as (see Appendix~\ref{sec:appendixB},~\eqref{phi0next})
\begin{equation}
    \phi_I = M_P\sqrt{\frac{\alpha}{\xi}}\,\text{arccoth}\left[\frac{\xi_h}{\sqrt{\alpha\xi}}\left(\displaystyle\sqrt{\frac{\alpha\xi}{\xi_h^2} + 2} + \sqrt{2}\right)\exp\left(\frac{4N_e\xi_h}{\alpha}\right)\right].
\end{equation}
Substituting this angle into~\eqref{faI} and then calculating the ratio~\eqref{eq:fa-ratio} gives
\begin{equation}\label{eq:accurate-ratio}
    \frac{f_{a,I}}{f_{a,0}} \simeq \sqrt{\frac{\alpha^2\xi\xi_\chi e^{\frac{8\xi_h N_e}{\alpha}}}{2\xi_h^2(\xi - \xi_h)\left(\sqrt{\frac{2\alpha\xi}{\xi_h^2} + 4} - 2\right) + \alpha\xi\left(\xi_h + \xi\left(e^{\frac{8\xi_h N_e}{\alpha}} - 1\right)\right)}}.
\end{equation}
This ratio is constrained by the CMB isocurvature bound.

The equation~~(\ref{eq:DM}) combined with~\eqref{eq:fa0} gives us the relation between the parameters $\gamma$ and $\xi_h$
\begin{equation}\label{eq:gamma}
    \gamma \sqrt{\xi_\chi} \simeq \left(\frac{5.6 \cdot 10^7 \langle r^2 \rangle}{R_a}\right)^{6/7} \sim 4.4 \cdot 10^6.
\end{equation}
This is the only relation on $\gamma$, since it cancels out from the amplitude of the isocurvature mode, see~\eqref{eq:isocurvature} and~\eqref{eq:fa-ratio}.

\begin{figure}[h!]
\begin{minipage}[h]{0.32\linewidth}
\center{\includegraphics[width=1\linewidth]{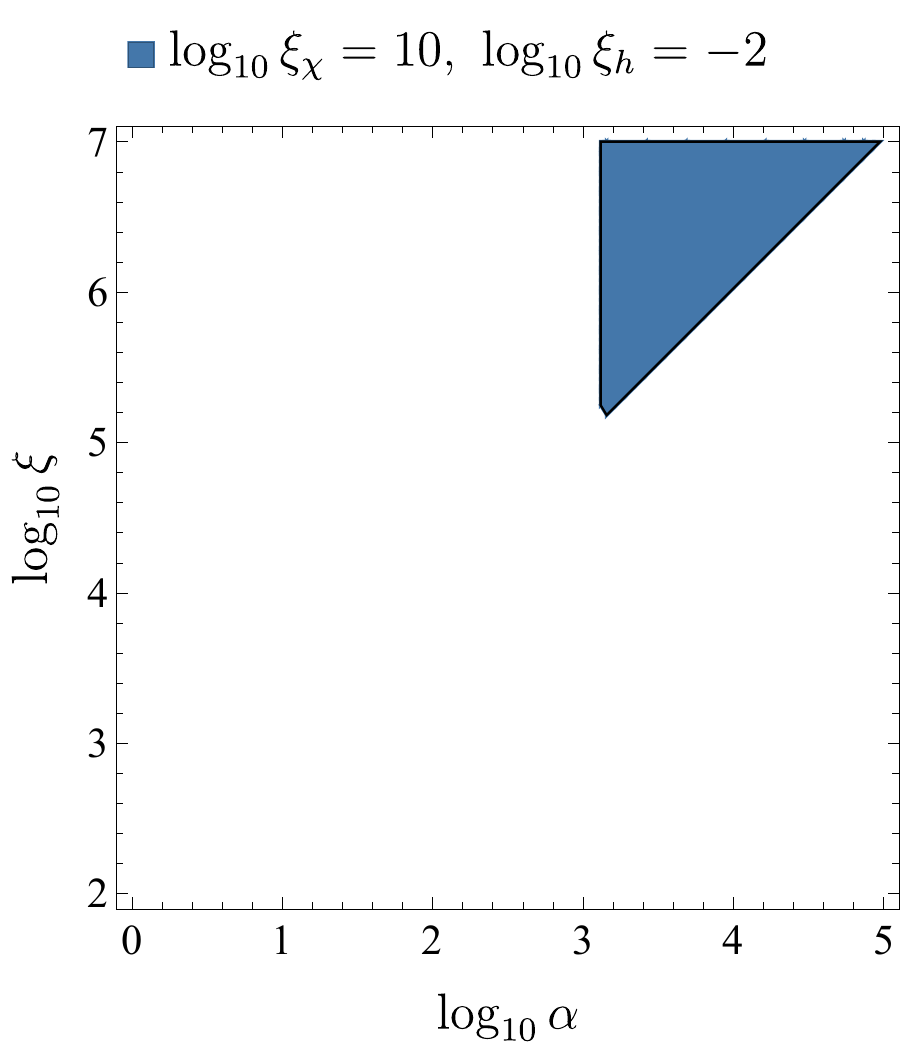}} \\
\end{minipage}
\begin{minipage}[h]{0.32\linewidth}
\center{\includegraphics[width=1\linewidth]{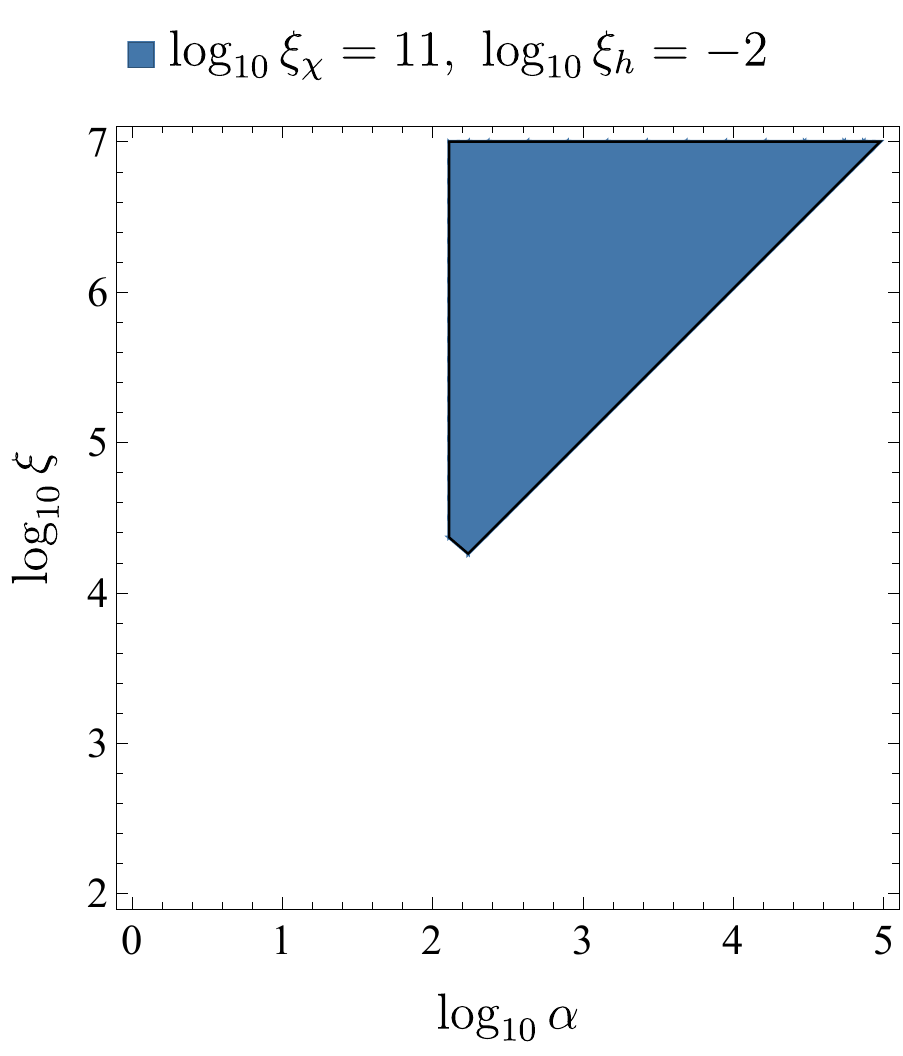}} \\
\end{minipage}
\begin{minipage}[h]{0.32\linewidth}
\center{\includegraphics[width=1\linewidth]{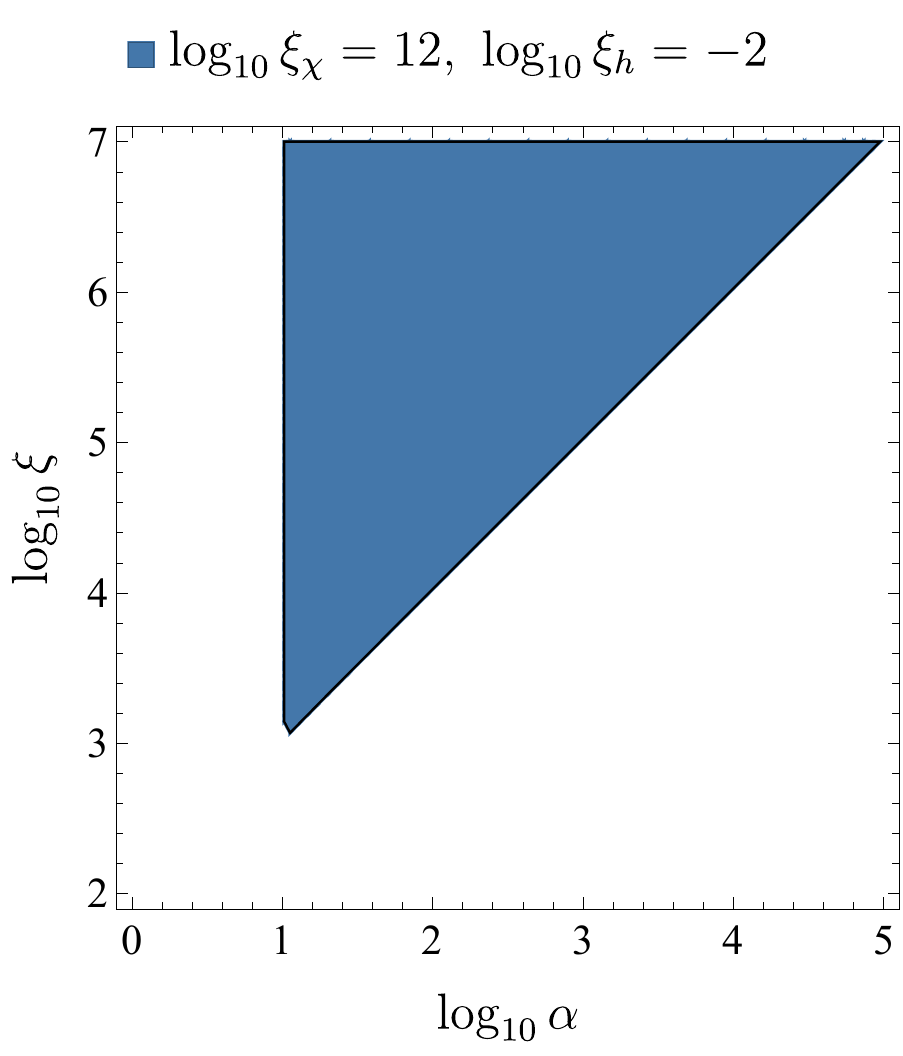}} \\
\end{minipage}
\vfill
\begin{minipage}[h]{0.32\linewidth}
\center{\includegraphics[width=1\linewidth]{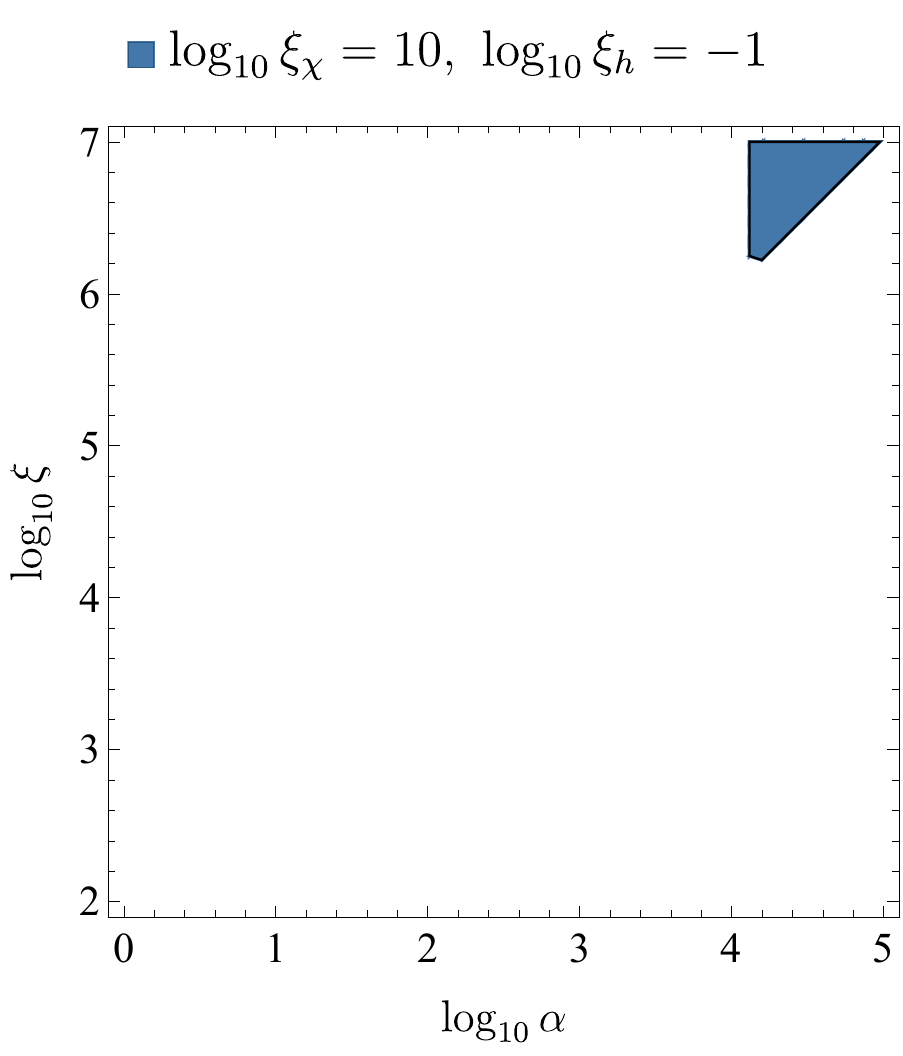}} \\
\end{minipage}
\begin{minipage}[h]{0.32\linewidth}
\center{\includegraphics[width=1\linewidth]{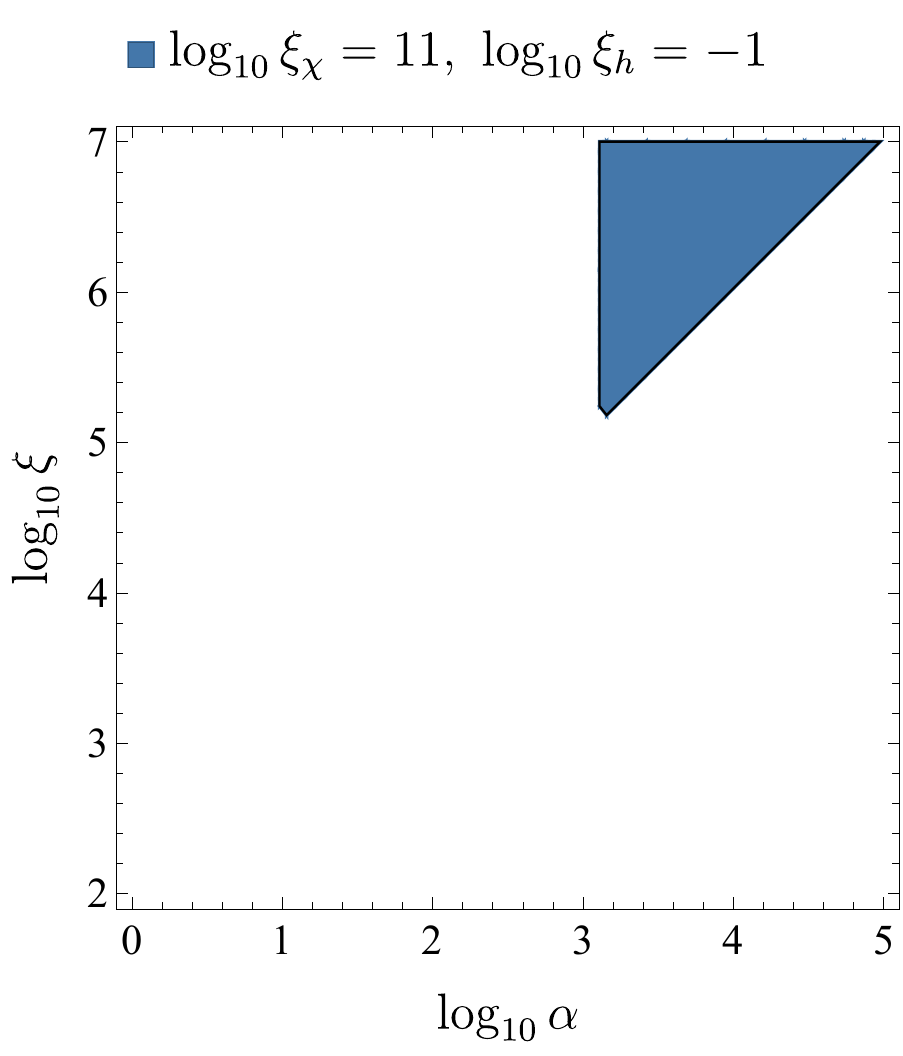}} \\
\end{minipage}
\begin{minipage}[h]{0.32\linewidth}
\center{\includegraphics[width=1\linewidth]{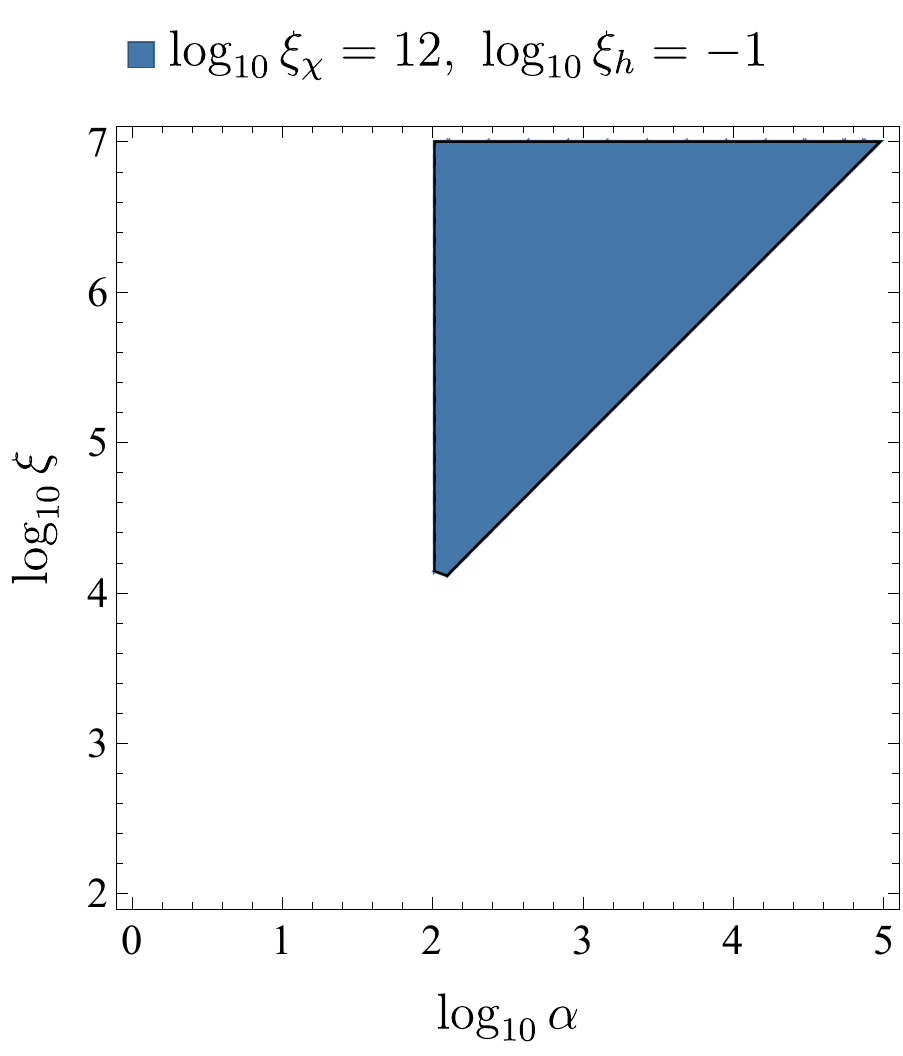}} \\
\end{minipage}
\caption{Cross-constraint on the parameters $\xi$, $\xi_\chi$, $\xi_h$ and $\alpha$ in logarithmic scale from observational bounds on $n_s$, $r_t$ and $\alpha_\text{iso}$. Here it is assumed that dark matter fully consists of the axion-dilaton condensate, and the tensor-to-scalar ratio $r_t$ is additionally suppressed by the requirement of unitarity~\eqref{eq:fa0-cutoff}.}
\label{fig:full-constraint}
\end{figure}

The isocurvature parameter of the axion is defined through the following ratio
\begin{equation}\label{eq:iso}
    \frac{\alpha_\text{iso}}{1 - \alpha_\text{iso}} \simeq \frac{1}{A_s}\frac{4R_a^2}{f_{a,I}^2\langle r^2 \rangle}\left(\frac{H_I}{2\pi}\right)^2,
\end{equation}
where the Hubble scale at inflation $H_I$ and the axion abundance in dark matter $R_a$ are given by~\eqref{eq:low-Hubble} and~\eqref{eq:DM}, respectively. Taking into account the stringent bound $\alpha_\text{iso} < \alpha_\text{iso,max} = 0.0019$ at the $95\%$ confidence level~\cite{Planck:2018vyg}, usually $f_{a,I} \gg f_{a,0}$ is required to avoid the isocurvature tension. The Planck constraint~\eqref{eq:iso} implies~\cite{Kearney:2016vqw}
\begin{equation}
    \frac{f_{a,I}}{f_{a,0}} \gtrsim 1.4 \cdot 10^4 \sqrt{R_a\left(\frac{r_t}{0.002}\right)\left(\frac{0.0019}{\alpha_\text{iso,max}}\right)}\left(\frac{0.1M_P}{f_{a,0}}\right)^{5/12}.
    \label{eq:Pierce-constraint}
\end{equation}
Substitution of~\eqref{eq:accurate-ratio} into~\eqref{eq:Pierce-constraint} gives the constraint on the parameters $\xi$, $\xi_h$, $\xi_\chi$ and $\alpha$ from the isocurvature bound. The final cross-constraint on the parameters of the model under the requirements that: (i)~axion-dilaton makes up all dark matter, while its initial displacement $r_\text{init}$ is not fine-tuned, i.e. $R_a \sim 1$ and $\langle r^2 \rangle \sim 1$ in~\eqref{eq:DM}; (ii)~inflation happens at energies lower than the cutoff scale~\eqref{eq:fa0-cutoff}; (iii)~the inflationary spectrum meets the observational constraints (see Fig.~\ref{fig:inflation-constraints}); (iv)~the ratio of the axion normalization constants in the Higgs vacuum and at inflation $f_{a,I}/f_{a,0}$ satisfies the isocurvature bound~\eqref{eq:Pierce-constraint} --- is shown in Fig.~\ref{fig:full-constraint}.


\section{Conclusions}\label{sec:VI}

We have presented a model, in which the spontaneously broken quantum scale invariance naturally works for solving the strong CP-problem in QCD under the assumption that the dilaton is coupled to the topological QCD term. At low energies, the model looks similar to the commonly discussed axion models, which opens the possibility to make the dilaton play the role of cold dark matter. As the universe cools down, the dilaton field (indentified with the QCD axion) acquires a mass term due to the non-perturbative QCD effects. This means that the scale symmetry, being the exact quantum symmetry of the model, is broken non-perturbatively.

The presented model is a low-energy EFT below the scale  of the QCD axion ($f_a\sim 10^{12}$ GeV), which generates dark matter without fine-tuning the initial dilaton field value, while at higher energies, the scale symmetry is restored in some strongly coupled UV completion, which is possibly related to quantum gravity effects. We have explored the possibility that all the phenomena in the early and late Universe are within the regime of validity of the effective description. In particular, we require the lower scale of inflation $H_I < f_a$, which at the same time helps to suppress the dilaton isocurvature perturbations. This lower scale $H_I$ is achieved by conside\-ring the dilaton dark matter scenario in a context of Higgs-driven inflation in Palatini formalism. We have shown that it is possible to meet the dark matter isocurvature constraint if the normalization constant of the dilaton kinetic term during inflation differs from that in the SM vacuum: $f_{a,I} \neq f_{a,0}$. We arrange this hierarchy and find the parameter space, which allows to meet all the CMB constraints on inflation.


\section*{Acknowledgements}
We thank Mikhail Shaposhnikov for helpful discussions. AB thanks Vasilii V. Pushkarev for careful reading and useful comments on earlier versions of the text. The work of AB is supported by the Russian Science Foundation grant 19-12-00393. AT is supported by Simons Foundation Award ID 555326 under the Simons Foundation Origins of the Universe initiative, Cosmology Beyond Einstein’s~Theory.\\


\appendix

\section{Derivation of the Einstein-frame Lagrangian}\label{sec:appendixA}

Here we present a detailed derivation of the Einstein-frame action~\eqref{eq:Ldiag}.

Let us write down again the general form of the Jordan-frame Lagrangian~\eqref{eq:SJ}
\begin{equation}
    \begin{aligned}
        &\mathcal{L}_J=-\frac{R}{2}\left(\xi_\chi \chi^2+\xi_h h^2+\xi\frac{h^4}{\chi^2}\right)+\frac{1}{2}(\partial\chi)^2\left(1+\alpha\frac{h^2}{\chi^2}\right)+\frac{1}{2}(\partial h)^2\left(1+\beta\frac{h^2}{\chi^2}\right)\\
        &-\frac{\lambda}{4}\left(h^2-\delta_0\chi^2\right)^2\left(1+\delta_1\frac{h^2}{\chi^2}+\delta_2\frac{h^4}{\chi^4}\right)-\frac{1}{4g^2}G^{\mu\nu}G_{\mu\nu}+\frac{\gamma}{64\pi^2}\log\left[\frac{\chi^2+h^2}{M_P^2}\right]\tilde{G}^{\mu\nu}G_{\mu\nu}.
    \end{aligned}
\end{equation}

We transform this Lagrangian to the Einstein frame as $g_{\mu\nu}\rightarrow e^{2\varrho} g_{\mu\nu}$, where
\begin{equation}\label{eq:Weyl_app}
    e^{2\varrho} = \frac{\xi_h h^2\chi^2 + \xi h^4 + \xi_\chi\chi^4}{\chi^2 M_P^2},
\end{equation}
is the Weyl factor. In Palatini formalism, the transformation of the Ricci scalar does not contribute to the kinetic terms of the fields. The relation between the quantities in the Einstein and Jordan frames are as follows
\begin{equation}\label{conf_transform}
    \hat{g}^{\mu\nu}=e^{-2\varrho}g^{\mu\nu}, \qquad \sqrt{|\hat{g}|}=e^{4\varrho}\sqrt{|g|}, \qquad \hat{R}\sqrt{|\hat{g}|}=e^{2\varrho}R\sqrt{|g|}.
\end{equation}
Hats denote the corresponding variables in the Einstein frame.

Therefore, the theory in the Einstein frame is described by the action
\begin{equation}
    \begin{aligned}
        &S_E=\int d^4x\sqrt{|\hat{g}|}\left[-\frac{\hat{R}}{2}M_P^2+\frac{1}{2}M_P^2\left(\partial\chi\right)^2\frac{\displaystyle\left(1+\alpha\frac{h^2}{\chi^2}\right)}{\displaystyle\left(\xi_h h^2+\xi\frac{h^4}{\chi^2}+\xi_\chi \chi^2\right)} + \frac{1}{2}M_P^2\left(\partial h\right)^2\frac{\displaystyle\left(1+\beta\frac{h^2}{\chi^2}\right)}{\displaystyle\left(\xi_h h^2+\xi\frac{h^4}{\chi^2}+\xi_\chi \chi^2\right)}\right.\\
        &\left. -\frac{\lambda M_P^4}{4}\,\frac{\displaystyle h^4\left(1+\delta_1\frac{h^2}{\chi^2}+\delta_2\frac{h^4}{\chi^4}\right)}{\displaystyle\left(\xi_h h^2+\xi\frac{h^4}{\chi^2}+\xi_\chi \chi^2\right)^2} - \frac{1}{4g^2}G^{\mu\nu}G_{\mu\nu}-\frac{\gamma}{64\pi^2}\log\left[\frac{\chi^2+h^2}{M_P^2}\right]\tilde{G}^{\mu\nu}G_{\mu\nu}\right].
    \end{aligned}
     \label{einstein_lagrangian}
\end{equation}

Note that gravity is now decoupled from the scalar sector, and the Lagrangian of the gauge fields is not changed by the Weyl transformation. We have also neglected the vacuum expectation value of the Higgs field compared to the Planck scale $M_P$. 

Let us choose new variables $h=\rho\sin\varphi$ and $\chi=\rho\cos\varphi$ to make the dilaton and the inflaton distinguishable. In terms of $\rho$ and $\varphi$, the action~\eqref{einstein_lagrangian} reads
\begin{equation}
\label{1}
    \begin{aligned}
        &S_E=\int d^4x\sqrt{|\hat{g}|}\left[-\frac{\hat{R}}{2}M_P^2+\frac{1}{2}M_P^2\,(\partial\varphi)^2\,\frac{\sin^2\varphi\left(\alpha\tan^2\varphi+\beta+1\right)+\cos^2\varphi}{\sin^2\varphi\left(\xi\tan^2\varphi+\xi_h\right)+\xi_\chi\cos^2\varphi}\right.\\
        &\left.+\frac{(\beta-\alpha)}{M_P^2}(\partial\varphi)\left(\partial\log\left[\frac{\rho}{M_P}\right]\right)\frac{\sin^2\varphi\tan\varphi}{\sin^2\varphi\left(\xi\tan^2\varphi+\xi_h\right)+\xi_\chi\cos^2\varphi}\right.\\
        &\left.+\frac{1}{2}M_P^2\left(\partial\log\left[\frac{\rho}{M_P}\right]\right)^2\frac{\sin^2\varphi\left(\alpha+\beta\tan^2\varphi+1\right)+\cos^2\varphi}{\sin^2\varphi\left(\xi\tan^2\varphi+\xi_h\right)+\xi_\chi\cos^2\varphi}\right.\\
        &-\left.\frac{\lambda M_P^4}{4}\frac{\sin^4\varphi\left(\delta_1\tan^2\varphi+\delta_2\tan^4\varphi+1\right)}{\left(\sin^2\varphi\left(\xi\tan^2\varphi+\xi_h\right)+\xi_\chi\cos^2\varphi\right)^2}-\frac{1}{4g^2}G^{\mu\nu}G_{\mu\nu}+\frac{\gamma}{32\pi^2}\log\left[\frac{\rho}{M_P}\right]\tilde{G}^{\mu\nu}G_{\mu\nu}\right].
    \end{aligned}
\end{equation}
In the Einstein frame, the potential does not depend on $\rho$, which means that it is a massless Goldstone mode. The corresponding symmetry, which $\rho$ belongs to, is scale invariance, which was explicitly broken by the Planck mass $M_P$ after performing the Weyl transformation~\eqref{eq:Weyl_app}--\eqref{conf_transform}.

To diagonalize the Lagrangian~\eqref{1}, we introduce a canonically normalized inflaton~$\phi$ and the dilaton~$r$ as follows\footnote{In order to derive such a change of variables, let us consider a non-diagonal kinetic part of a Lagrangian of two fields $\rho$ and $\varphi$ of the form
\begin{equation}\label{eq:non-diag-footnote-1}
    \mathcal{L} \ni f_1^2(\varphi)\left(\partial\ln\rho\right)^2 + f_2^2(\varphi)\left(\partial\ln\rho\right)\left(\partial\varphi\right) + f_3^2(\varphi)\left(\partial\varphi\right)^2,
\end{equation}
where we assumed that the coefficients $f_{1,2,3}$ depend only on $\varphi$, as in~\eqref{1}. In our context, the field $\varphi$ is the Higgs-inflaton, so in order to study the spectrum of inflation, we need to canonically normalize the kinetic term of $\varphi$ as well as diagonalize~\eqref{eq:non-diag-footnote-1}. Denoting $F(\varphi) = \displaystyle\frac{f_2^2(\varphi)}{f_1^2(\varphi)}$ and $A(\varphi) = \displaystyle\frac{f_3^2(\varphi)}{f_1^2(\varphi)}$, we obtain
\begin{equation}\label{eq:non-diag-footnote-2}
    \mathcal{L} \ni f_1^2(\varphi)\left[\left(\partial\ln\rho\right)^2 + A(\varphi)\left(\partial\varphi\right)^2 + F(\varphi)\left(\partial\ln\rho\right)\left(\partial\varphi\right)\right].
\end{equation}
Therefore, introducing a new field
\begin{equation}
    \ln r = \ln\rho + \frac{1}{2}\int\limits^\varphi_0 F(s)ds,
\end{equation}
we diagonalize~\eqref{eq:non-diag-footnote-2}
\begin{equation}
    \mathcal{L} \ni f_1^2(\varphi)\left[\left(\partial\ln r\right)^2 + \left(A(\varphi) - \frac{1}{4}F^2(\varphi)\right)\left(\partial\varphi\right)^2\right].
\end{equation}
From the condition
\begin{equation}
    f_1^2(\varphi)\left(A(\varphi) - \frac{1}{4}F^2(\varphi)\right)\left(\partial\varphi\right)^2 \equiv \frac{1}{2}\left(\partial\phi\right)^2,
\end{equation}
we obtain the Jacobian for the canonically normalized inflaton
\begin{equation}
    \phi = \int\limits^\varphi_0 ds \sqrt{2 f_1^2(\varphi)\left(A(\varphi) - \frac{1}{4}F^2(\varphi)\right)}.
\end{equation}
}
\begin{equation}
\label{2}
    \phi(\varphi)=\frac{M_P^2}{2}\int\limits^\varphi_0 ds\,\sqrt{\frac{\sec^8 s\Big((\alpha -1)\cos 2s-\alpha-1\Big)\Big((\beta-1)\cos 2s-\beta-1\Big)}{\Big((\alpha+1)\tan^2s+\beta\tan^4s+1\Big)\Big(\xi\tan^4s+\xi_h\tan^2s+\xi_\chi\Big)}},
\end{equation}
\begin{equation}
\label{2.1}
        r=\log\left[\frac{\rho}{M_P}\right] + (\beta-\alpha)\int\limits^\varphi_0 ds\,\frac{\sin^2s\tan s}{\sin^2s\left(\alpha+\beta\tan^2s+1\right)+\cos^2s}\equiv\log\left[\frac{\rho}{M_P}\right]+F(\varphi).
\end{equation}
Note that for the case $\alpha=\beta$, the kinetic term in~\eqref{1} diagonalizes automatically. 

After such a change of variables, the Lagrangian (\ref{1}) diagonalizes
\begin{equation}\label{3}
    \begin{aligned}
         &S_E=\int d^4x\sqrt{|\hat{g}|}\left[-\frac{\hat{R}}{2}M_P^2+\frac{1}{2}M_P^2\,\frac{\sin^2\varphi(\phi)\left(\alpha+\beta\tan^2\varphi(\phi)+1\right)+\cos^2\varphi(\phi)}{\sin^2\varphi(\phi)\left(\xi\tan^2\varphi(\phi)+\xi_h\right)+\xi_\chi\cos^2\varphi(\phi)}\left(\partial r\right)^2\right.\\
         &\left.+\frac{1}{2}\left(\partial\phi\right)^2-\frac{\lambda M_P^4}{4}\,\frac{\sin^4\varphi(\phi)\Big(\delta_1\tan^2\varphi(\phi)+\delta_2\tan^4\varphi(\phi)+1\Big)}{\Big(\sin^2\varphi(\phi)\left(\xi\tan^2\varphi(\phi)+\xi_h\right)+\xi_\chi\cos^2\varphi(\phi)\Big)^2}\right.\\
         &-\left.\frac{1}{4g^2}G^{\mu\nu}G_{\mu\nu}-\frac{\gamma}{32\pi^2}\Big(r-F\left(\varphi(\phi)\right)\Big)\tilde{G}^{\mu\nu}G_{\mu\nu}\right],
    \end{aligned}
\end{equation}
where $\varphi(\phi)$ is given by the inverse of (\ref{2}).

Phenomenologically, the most intriguing case is $\alpha=\beta$. In this case, $F(\varphi(\phi))$ in~\eqref{2.1} vanishes, and the QCD part of the action~\eqref{3} resembles that of the QCD axion models.


\section{Analytical formulas for primordial spectrum}\label{sec:appendixB}

Here we study inflation described by the Lagrangian~\eqref{einstein_lagrangian} with $\alpha=\beta$.

\subsection*{Low-energy limit}

First, let us convince ourselves that the solution of~\eqref{2} gives the standard Higgs potential in the low-energy limit. We start directly from~\eqref{2} with $\alpha=\beta$
\begin{equation}\label{phi}
    \phi(\varphi) = \frac{M_P}{2}\int\limits^\varphi_0 ds\,\frac{|\sec s|^3\Big(\alpha + 1 - (\alpha - 1)\cos 2s\Big)}{\sqrt{\Big(\alpha\tan^2 s + 1\Big)\Big(\xi\tan^4 s + \xi_h\tan^2 s + \xi_\chi\Big)}}.
\end{equation}

The potential in~\eqref{3} has a minimum at $\varphi=0$. Expanding~\eqref{phi} around this minimum, we obtain
\begin{equation}\label{eq:phi-approx}
    \phi(\varphi)=\frac{M_P}{\sqrt{\xi_\chi}}\,\varphi + O\left(\varphi^3\right).
\end{equation}

Then, at first order
\begin{equation}
    \varphi(\phi)=\sqrt{\xi_\chi}\frac{\phi}{M_P}.
\end{equation}
We see that the potential in terms of the canonically normalized Higgs $\phi$ also has a minimum at $\phi = 0$. Therefore, we obtain the low-energy potential 
\begin{equation}
    V(\varphi)\simeq\frac{\lambda M_P^4}{4\xi_\chi^2}\,\varphi^4+O\left(\varphi ^5\right),
\end{equation}
which in terms of $\phi$~\eqref{eq:phi-approx} becomes the SM Higgs potential
\begin{equation}
    V(\phi)\simeq\frac{\lambda\phi^4}{4}.
\end{equation}

\subsection*{Effective Lagrangian for inflation}

The general solution of~\eqref{phi} is impossible to find analytically for an arbitrary value of~$\varphi$, so now we turn our attention back to~\eqref{einstein_lagrangian}. Since $\langle\chi\rangle\simeq M_P$, the inflation regime $h \gg M_P$ is equivalent to $h \gg \chi$. In this limit, the Lagrangian (\ref{einstein_lagrangian}) takes the form
\begin{equation}
    \mathcal{L}_{E,\,I} \simeq -\frac{1}{2}M_P^2 R  +\frac{1}{2}\alpha M_P^2\,\frac{(\partial h)^2 + (\partial\chi)^2}{\xi_h\chi^2 + \xi h^2} - \frac{\lambda M_P^4}{4}\frac{\delta_2 h^4}{\left(\xi_h\chi^2 + \xi h^2\right)^2}.
\end{equation}

The potential at super-Planckian scales is determined mostly by the parameters $\delta_2$, $\xi$ and $\xi_h$, while other parameters are the subleading corrections suppressed in the limit $h\gg M_P$. After choosing the same field variables $\varphi$ and $r$ (see~\eqref{2.1} with $\alpha=\beta$), we arrive at
\begin{equation}\label{inflation}
    \mathcal{L}_{E,I} = -\frac{1}{2}M_P^2 R + \frac{1}{2}\alpha M_P^2\,\frac{(\partial r)^2 + (\partial\varphi)^2}{\xi_h\cos^2\varphi + \xi\sin^2\varphi} - \frac{\lambda\delta_2 M_P^4}{4}\frac{\sin^4\varphi}{\left(\xi_h\cos^2\varphi + \xi\sin^2\varphi\right)^2}.
\end{equation}

Canonical normalization of the Higgs field $\varphi$ leads to
\begin{equation}
    \varphi=\text{am}\left(\sqrt{\frac{\xi_h}{\alpha}}\frac{\phi-\phi_\text{vac}}{M_P},1-\frac{\xi }{\xi_h}\right),
\end{equation}
where am$(x,y)$ is the Jacobi amplitude, and the integration constant $\phi_\text{vac}$ is to be chosen so that the potential in~\eqref{inflation} has a minimum at $\varphi=0$.

On the plateau of the potential in~\eqref{inflation}, in which we can neglect the first term in the denominator compared to the second one, we approximately have
\begin{equation}\label{canon_field}
    \tan\frac{\varphi}{2}=\exp\left[\sqrt{\frac{\xi}{\alpha}}\frac{\phi-\phi_\text{vac}}{M_P}\right].
\end{equation}

The approximate inflationary potential in terms of the canonical inflaton $\phi$ is given by the substitution of~\eqref{canon_field} back to~\eqref{inflation}
\begin{equation}
    V_{E}(\phi) \simeq \frac{\lambda\delta_2 M_P^4}{4\xi^2} \left(1 + \frac{\xi_h}{\xi}\sinh^2\left[\sqrt{\frac{\xi}{\alpha}}\frac{\phi - \phi_\text{vac}}{M_P}\right]\right)^{-2}.
\end{equation}

Near the plateau, we get
\begin{equation}\label{eq:Inflationary-pot}
    V_E(\phi) \simeq \frac{\lambda\delta_2 M_P^4}{4\xi^2}\left(1 - \frac{\xi_h}{\xi}\sinh^2\left[\sqrt{\frac{\xi}{\alpha}}\frac{\phi - \phi_\text{vac}}{M_P}\right]\right)^{2}.
\end{equation}

\subsection*{Primordial spectrum}

Let us calculate the primordial spectrum generated by the potential~\eqref{eq:Inflationary-pot}.

The number of e-folds is defined as
\begin{equation}
    N_e=\frac{1}{M_P^2}\int\limits^{\phi_I}_{\phi_e}d\phi\,\frac{V_E(\phi)}{V_E^\prime(\phi)},
\end{equation}
where $\phi_I$ and $\phi_e$ are the initial and final values of the canonical field, respectively. For the potential~\eqref{eq:Inflationary-pot}, it is given by
\begin{equation}\label{Ne}
    N_e = \frac{\alpha}{4\xi\xi_h}\left((\xi + \xi_h)\ln\left[\frac{\displaystyle\cosh\sqrt{\frac{\xi}{\alpha}}\frac{\phi_I}{M_P}}{\displaystyle\cosh\sqrt{\frac{\xi}{\alpha}}\frac{\phi_e}{M_P}}\right] - \xi\ln\left[\frac{\displaystyle\sinh\sqrt{\frac{\xi}{\alpha}}\frac{\phi_I}{M_P}}{\displaystyle\sinh\sqrt{\frac{\xi}{\alpha}}\frac{\phi_e}{M_P}}\right]\right),
\end{equation}

The initial value $\phi_I$ during inflation in~\eqref{Ne} is to solve the horizon problem
\begin{equation}
    \phi_I: \quad N_e(\phi_I) \simeq 60.
\end{equation}

We are able to resolve~\eqref{Ne} with respect to $\phi_I$ under some natural assumptions. From the potential in~\eqref{inflation} we see that inflation is dominated by the mixed-field coupling to gravity $\xi$ in contrast to the standard Higgs or Higgs-dilaton scenarios, where inflation is determined by the Higgs-gravity coupling $\xi_h$. In our case, the Higgs is still the inflaton, but its coupling to gravity $\xi_h$ is subdominant. Therefore, we require the Higgs coupling $\xi_h$ be much smaller than $\xi$
\begin{equation}
    \xi \gg \xi_h.
\end{equation}
This fact reduces~\eqref{Ne} to the following equation
\begin{equation}
    N_e = \frac{\alpha}{4\xi_h}\ln\left[\coth\left(\sqrt{\frac{\xi}{\alpha}}\frac{\phi_I}{M_P}\right)\tanh\left(\sqrt{\frac{\xi}{\alpha}}\frac{\phi_e}{M_P}\right)\right],
\end{equation}
which has a solution
\begin{equation}
    \phi_I = \sqrt{\frac{\alpha}{\xi}}M_P\,\text{arccoth}\left[\exp\left(\frac{4\xi_h N_e}{\alpha}\right)\coth\left(\sqrt{\frac{\xi}{\alpha}}\frac{\phi_e}{M_P}\right)\right],
    \label{phi0}
\end{equation}
where $\phi_e$ is the value of the Higgs when the slow-roll regime ends. To find the expression for $\phi_e$, let us calculate the slow-roll parameter $\epsilon$ for the potential~\eqref{eq:Inflationary-pot}
\begin{equation}
    \epsilon \simeq \frac{2\xi_h^2}{\alpha\xi}\sinh^2\left(2\sqrt{\frac{\xi}{\alpha}}\frac{\phi}{M_P}\right), \qquad \xi \gg \xi_h.
    \label{epsapprox}
\end{equation}

The slow-roll regime ends when $\epsilon\simeq 1$
\begin{equation}
    \phi_e: \quad \epsilon(\phi_e) \simeq 1,
\end{equation}
which defines the final value $\phi_e$ of the field $\phi$. From~\eqref{epsapprox} we find
\begin{equation}
    \phi_e=\frac{1}{2}M_P\sqrt{\frac{\alpha}{\xi}}\,\text{arcsinh}\left(\sqrt{\frac{\alpha\xi}{2}}\frac{1}{\xi_h}\right).
\end{equation}

Substituting this expression back into~\eqref{phi0}, we express the initial value $\phi_I$ in terms of the parameters $\xi$, $\xi_h$, $\alpha$ and the number of e-folds $N_e$
\begin{equation}\label{phi0next}
    \phi_I = M_P\sqrt{\frac{\alpha}{\xi}}\,\text{arccoth}\left[\frac{\xi_h}{\sqrt{\alpha\xi}}\left(\displaystyle\sqrt{\frac{\alpha\xi}{\xi_h^2} + 2} + \sqrt{2}\right)\exp\left(\frac{4N_e\xi_h}{\alpha}\right)\right].
\end{equation}

In the end, we derive the expression for the approximate slow-roll parameter $\epsilon$ as a function of e-folds $N_e$ by substituting~\eqref{phi0next} into~\eqref{epsapprox}
\begin{equation}\label{eq:eps-Ne}
    \epsilon \simeq \frac{4\xi_h^2}{2\xi_h^2\displaystyle\sqrt{\frac{2\alpha\xi}{\xi_h^2} + 4}\,\sinh\left(\displaystyle\frac{8N_e\xi_h}{\alpha}\right) + \left(\alpha\xi + 4\xi_h^2\right)\cosh\left(\displaystyle\frac{8N_e\xi_h}{\alpha}\right) - \alpha\xi}.
\end{equation}
Taking into account that $\alpha,\xi \gg \xi_h$ (see Fig.~\ref{fig:inflation-constraints} and~\ref{fig:full-constraint}), the expression~\eqref{eq:eps-Ne} simplifies~to
\begin{equation}\label{eq:eps-simple}
    \epsilon \simeq \frac{\alpha}{4N_e\left(2\xi N_e + \sqrt{2\alpha\xi}\right)}.
\end{equation}

The same procedure gives us the second slow-roll parameter $\eta$ as a function of $N_e$
\begin{equation}\label{eq:eta-Ne}
    \eta \simeq \frac{8\xi_h^3\displaystyle\sqrt{\frac{2\alpha\xi}{\xi_h^2} + 4} - 4\alpha\xi_h\xi\sinh\left(\displaystyle\frac{8N_e\xi_h}{\alpha}\right)}{\alpha\left(-\alpha\xi + \alpha\xi\cosh\left(\displaystyle\frac{8N_e\xi_h}{\alpha}\right) - 4\xi_h^2\right)},
\end{equation}
which simplifies to
\begin{equation}\label{eq:eta-simple}
    \eta \simeq -\frac{4\xi}{4\xi N_e + \sqrt{2\alpha\xi}}.
\end{equation}
In numerical calculations, we use~\eqref{eq:eps-Ne} and~\eqref{eq:eta-Ne} to probe the influence of the parameter $\xi_h$, which is subdominant compared to $\alpha$ and $\xi$. However, the general behaviour of the inflationary spectra can be estimated with the use of~\eqref{eq:eps-simple} and~\eqref{eq:eta-simple}, which in fact depend only on the ratio $\alpha/\xi$.

The results~\eqref{eq:eps-Ne} and~\eqref{eq:eta-Ne} allow us to calculate the analytical formulas for the primordial spectrum of the inflation governed by the potential~\eqref{eq:Inflationary-pot}:
\begin{itemize}
    \item The spectral tilt $n_s$
        \begin{equation}
            \begin{aligned}
            n_s \simeq 1 - \frac{3\alpha}{2N_e\left(2\xi N_e + \sqrt{2\alpha\xi}\right)} - \frac{8\xi}{4\xi N_e + \sqrt{2\alpha\xi}}.
            \end{aligned}
        \end{equation}
    \item Tensor-to-scalar ratio $r_t$
            \begin{equation}
            r_t \simeq \frac{4\alpha}{N_e\left(2\xi N_e + \sqrt{2\alpha\xi}\right)}.
        \end{equation}
    \item The amplitude of scalar perturbations $A_s$
        \begin{equation}\label{eq:As}
            \begin{aligned}
            &A_s \simeq \frac{\lambda\delta_2}{1536\pi^2\xi^4\alpha N_e}(\alpha - 8\xi N_e)^2\left(2\xi N_e + \sqrt{2\alpha\xi}\right).
            \end{aligned}
        \end{equation}
The constraint~\eqref{eq:As-Planck} on the amplitude $A_s$ fixes the value of $\delta_2$, i.e. the height of the potential~\eqref{eq:Inflationary-pot}. 
\end{itemize}

\section{Why not simpler?}\label{sec:appendixC}

At first sight, the Lagrangian~\eqref{eq:SJ} does not seem the most natural because of the non-trivial scalar-gravity coupling~\eqref{eq:Lgrav}, the corrections to the kinetic terms of the fields~\eqref{eq:KHD} and to the potential~\eqref{eq:VHD}. Here we explain why~\eqref{eq:SJ} is in fact the most simple model for the scale invariant axion consistent with observational constraints.\\

Let us start with a simpler Lagrangian with the same axion-dilaton field
\begin{equation}
    \frac{\mathcal{L}_J}{\sqrt{-g}} = -\frac{R}{2}\left(\xi_h h^2 + \xi_\chi\chi^2\right)^2 + \frac{1}{2}(\partial\chi)^2 + \frac{1}{2}(\partial h)^2 - \frac{\lambda}{4}\left(h^2 - \delta_0\chi^2\right)^2 - \mathcal{L_\text{QCD}},
\end{equation}
where
\begin{equation}
    \mathcal{L_\text{QCD}} = -\frac{1}{4g^2}G^{\mu\nu}G_{\mu\nu} + \frac{\gamma}{64\pi^2}\log\left[\frac{\chi^2 + h^2}{M_P^2}\right]\tilde{G}^{\mu\nu}G_{\mu\nu}.
\end{equation}

After the Weyl transformation with the factor $e^{2\varrho}$ given by
\begin{equation}
    e^{2\varrho} = \frac{\xi_h h^2 + \xi_\chi \chi^2}{M_P^2},
\end{equation}
we obtain the Einstein-frame Lagrangian
\begin{equation}
    \frac{\mathcal{L}_E}{\sqrt{-\hat{g}}} = -\frac{1}{2}M_P^2 R + \frac{1}{2}M_P^2\frac{(\partial h)^2 + (\partial \chi)^2}{\xi_h h^2 + \xi_\chi \chi^2} - \frac{\lambda M_P^4}{4}\frac{(h^2 - \delta_0 \chi^2)^2}{(\xi_h h^2 + \xi_\chi \chi^2)^2} + \mathcal{L}_\text{QCD},
\end{equation}
where $\mathcal{L}_\text{QCD}$ gets unchanged under the Weyl transformation.

Introducing new fields in full analogy with the previous sections
\begin{equation}
    \begin{aligned}
        &r = \log\frac{\sqrt{h^2 + \chi^2}}{M_P}, \qquad &\text{axion-dilaton}\\
        &\varphi = \arctan\frac{h}{\chi}, \qquad &\text{inflaton},
    \end{aligned}
\end{equation}
we obtain the final Langrangian
\begin{equation}
\begin{aligned}
    \frac{\mathcal{L}_E}{\sqrt{-\hat{g}}} &= -\frac{1}{2}M_P^2\hat{R} + \frac{1}{2}M_P^2\frac{(\partial r)^2 + (\partial\varphi)^2}{\xi_h \sin^2\varphi + \xi_\chi\cos^2\varphi}\\
    & - \frac{\lambda M_P^4}{4}\frac{\sin^4\varphi}{(\xi_h \sin^2\varphi + \xi_\chi \cos^2\varphi)^2} - \frac{1}{4g^2}G^{\mu\nu}G_{\mu\nu} + \frac{\gamma}{32\pi^2}\,r\,\tilde{G}^{\mu\nu}G_{\mu\nu}.
    \end{aligned}
    \label{eq:Simple-L-EF}
\end{equation}
One can see an important difference between~\eqref{eq:Simple-L-EF} and~\eqref{eq:Ldiag}: in the initial model~\eqref{eq:SJ}, inflation does not depend on $\xi_\chi$, which allows to use the latter along with $\gamma$ to tune the amount of the axion abundance~\eqref{eq:DM} in dark matter, regardless of the constraints on the inflationary spectrum. In the model~\eqref{eq:Simple-L-EF}, $\xi_\chi$ is responsible for both the spectrum and the axion fraction, which runs the risk of being potentially inconsistent with cross-constraints. As we will see, this is exactly what happens.\\

The inflationary spectrum of the model~\eqref{eq:Simple-L-EF} is known~\cite{Garcia-Bellido:2011kqb,Rubio:2020zht,Casas:2017wjh} and can be obtained by following the same procedure described in Appendices~\ref{sec:appendixA} and~\ref{sec:appendixB}. We list the final results:
\begin{itemize}
    \item The slow-roll parameters are
    \begin{equation}
        \epsilon \simeq \frac{1}{2\xi_h N_e}, \qquad \eta \simeq -\frac{1}{N_e}.
    \end{equation}
    \item The spectral index is
        \begin{equation}
            n_s - 1 \simeq -\frac{2}{N_e}.
        \end{equation}
    \item The tensor-to-scalar ratio is given by
        \begin{equation}
            r_t \simeq \frac{8}{\xi_h N_e^2}.
        \end{equation}
    \item The amplitude of scalar perturbations reads
        \begin{equation}
            A_s \simeq \frac{\lambda N_e^2}{72\pi^2\xi_h^2}.
        \end{equation}
\end{itemize}

In order to satisfy the Planck normalization, we need
\begin{equation}
    \xi_h \simeq \frac{N_e\lambda^{1/2}}{6\pi\sqrt{2A_s}},
\end{equation}
Numerically, in the limit $\xi_h \gg 1$ and $N_e = 60$
\begin{equation}
    \epsilon \simeq \frac{2 \times 10^{-4}}{\xi_h}, \qquad
    \eta \simeq -0.02, \qquad
    \xi_h \simeq \lambda^{1/2} \times 10^9.
\end{equation}

The Hubble scale of this model is also relatively low
\begin{equation}
    \frac{H_I}{M_P} \sim \frac{\lambda^{1/2}}{\xi_h}.
\end{equation}

The axion constant as a function of the inflaton $\varphi$ is given by
\begin{equation}
    f_a(\varphi) = \frac{M_P}{\gamma\sqrt{\xi_h\sin^2\varphi + \xi_\chi\cos^2\varphi}}.
\end{equation}

Numerically, taking $\xi_h \sim 10^5$, $\xi_\chi \sim 10^{-3}$ and $N_e = 60$, which corresponds to ${\varphi \simeq 0.67}$, we get $f_{a,I}/f_a \sim 10^{-4}$, and together with~\eqref{eq:DM}, the isocurvature constraint~\eqref{eq:Pierce-constraint} implies that the axion displacement needs to be hugely fine-tuned to meet the observational constraints
\begin{equation}
    \langle r^2 \rangle \lesssim 10^{-22}.
\end{equation}

\bibliography{paper}

\begin{thebibliography}{52}%
\makeatletter
\providecommand \@ifxundefined [1]{%
 \@ifx{#1\undefined}
}%
\providecommand \@ifnum [1]{%
 \ifnum #1\expandafter \@firstoftwo
 \else \expandafter \@secondoftwo
 \fi
}%
\providecommand \@ifx [1]{%
 \ifx #1\expandafter \@firstoftwo
 \else \expandafter \@secondoftwo
 \fi
}%
\providecommand \natexlab [1]{#1}%
\providecommand \enquote  [1]{``#1''}%
\providecommand \bibnamefont  [1]{#1}%
\providecommand \bibfnamefont [1]{#1}%
\providecommand \citenamefont [1]{#1}%
\providecommand \href@noop [0]{\@secondoftwo}%
\providecommand \href [0]{\begingroup \@sanitize@url \@href}%
\providecommand \@href[1]{\@@startlink{#1}\@@href}%
\providecommand \@@href[1]{\endgroup#1\@@endlink}%
\providecommand \@sanitize@url [0]{\catcode `\\12\catcode `\$12\catcode
  `\&12\catcode `\#12\catcode `\^12\catcode `\_12\catcode `\%12\relax}%
\providecommand \@@startlink[1]{}%
\providecommand \@@endlink[0]{}%
\providecommand \url  [0]{\begingroup\@sanitize@url \@url }%
\providecommand \@url [1]{\endgroup\@href {#1}{\urlprefix }}%
\providecommand \urlprefix  [0]{URL }%
\providecommand \Eprint [0]{\href }%
\providecommand \doibase [0]{https://doi.org/}%
\providecommand \selectlanguage [0]{\@gobble}%
\providecommand \bibinfo  [0]{\@secondoftwo}%
\providecommand \bibfield  [0]{\@secondoftwo}%
\providecommand \translation [1]{[#1]}%
\providecommand \BibitemOpen [0]{}%
\providecommand \bibitemStop [0]{}%
\providecommand \bibitemNoStop [0]{.\EOS\space}%
\providecommand \EOS [0]{\spacefactor3000\relax}%
\providecommand \BibitemShut  [1]{\csname bibitem#1\endcsname}%
\let\auto@bib@innerbib\@empty
\bibitem [{\citenamefont {Wetterich}(1984)}]{WETTERICH1984215}%
  \BibitemOpen
  \bibfield  {author} {\bibinfo {author} {\bibfnamefont {C.}~\bibnamefont
  {Wetterich}},\ }\bibfield  {title} {\bibinfo {title} {Fine-tuning problem and
  the renormalization group},\ }\href
  {https://doi.org/https://doi.org/10.1016/0370-2693(84)90923-7} {\bibfield
  {journal} {\bibinfo  {journal} {Physics Letters B}\ }\textbf {\bibinfo
  {volume} {140}},\ \bibinfo {pages} {215} (\bibinfo {year}
  {1984})}\BibitemShut {NoStop}%
\bibitem [{\citenamefont {Bardeen}(1995)}]{Bardeen:1995kv}%
  \BibitemOpen
  \bibfield  {author} {\bibinfo {author} {\bibfnamefont {W.~A.}\ \bibnamefont
  {Bardeen}},\ }\bibfield  {title} {\bibinfo {title} {{On naturalness in the
  standard model}},\ }in\ \href@noop {} {\emph {\bibinfo {booktitle} {{Ontake
  Summer Institute on Particle Physics}}}}\ (\bibinfo {year}
  {1995})\BibitemShut {NoStop}%
\bibitem [{\citenamefont {Shaposhnikov}\ and\ \citenamefont
  {Shkerin}(2018{\natexlab{a}})}]{Shaposhnikov:2018xkv}%
  \BibitemOpen
  \bibfield  {author} {\bibinfo {author} {\bibfnamefont {M.}~\bibnamefont
  {Shaposhnikov}}\ and\ \bibinfo {author} {\bibfnamefont {A.}~\bibnamefont
  {Shkerin}},\ }\bibfield  {title} {\bibinfo {title} {{Conformal symmetry:
  towards the link between the Fermi and the Planck scales}},\ }\href
  {https://doi.org/10.1016/j.physletb.2018.06.068} {\bibfield  {journal}
  {\bibinfo  {journal} {Phys. Lett. B}\ }\textbf {\bibinfo {volume} {783}},\
  \bibinfo {pages} {253} (\bibinfo {year} {2018}{\natexlab{a}})},\ \Eprint
  {https://arxiv.org/abs/1803.08907} {arXiv:1803.08907 [hep-th]} \BibitemShut
  {NoStop}%
\bibitem [{\citenamefont {Shaposhnikov}\ and\ \citenamefont
  {Shkerin}(2018{\natexlab{b}})}]{Shaposhnikov:2018jag}%
  \BibitemOpen
  \bibfield  {author} {\bibinfo {author} {\bibfnamefont {M.}~\bibnamefont
  {Shaposhnikov}}\ and\ \bibinfo {author} {\bibfnamefont {A.}~\bibnamefont
  {Shkerin}},\ }\bibfield  {title} {\bibinfo {title} {{Gravity, Scale
  Invariance and the Hierarchy Problem}},\ }\href
  {https://doi.org/10.1007/JHEP10(2018)024} {\bibfield  {journal} {\bibinfo
  {journal} {JHEP}\ }\textbf {\bibinfo {volume} {10}},\ \bibinfo {pages}
  {024}},\ \Eprint {https://arxiv.org/abs/1804.06376} {arXiv:1804.06376
  [hep-th]} \BibitemShut {NoStop}%
\bibitem [{\citenamefont {Shaposhnikov}\ \emph {et~al.}(2021)\citenamefont
  {Shaposhnikov}, \citenamefont {Shkerin},\ and\ \citenamefont
  {Zell}}]{Shaposhnikov:2020geh}%
  \BibitemOpen
  \bibfield  {author} {\bibinfo {author} {\bibfnamefont {M.}~\bibnamefont
  {Shaposhnikov}}, \bibinfo {author} {\bibfnamefont {A.}~\bibnamefont
  {Shkerin}},\ and\ \bibinfo {author} {\bibfnamefont {S.}~\bibnamefont
  {Zell}},\ }\bibfield  {title} {\bibinfo {title} {{Standard Model Meets
  Gravity: Electroweak Symmetry Breaking and Inflation}},\ }\href
  {https://doi.org/10.1103/PhysRevD.103.033006} {\bibfield  {journal} {\bibinfo
   {journal} {Phys. Rev. D}\ }\textbf {\bibinfo {volume} {103}},\ \bibinfo
  {pages} {033006} (\bibinfo {year} {2021})},\ \Eprint
  {https://arxiv.org/abs/2001.09088} {arXiv:2001.09088 [hep-th]} \BibitemShut
  {NoStop}%
\bibitem [{\citenamefont {Englert}\ \emph {et~al.}(1976)\citenamefont
  {Englert}, \citenamefont {Truffin},\ and\ \citenamefont
  {Gastmans}}]{Englert:1976ep}%
  \BibitemOpen
  \bibfield  {author} {\bibinfo {author} {\bibfnamefont {F.}~\bibnamefont
  {Englert}}, \bibinfo {author} {\bibfnamefont {C.}~\bibnamefont {Truffin}},\
  and\ \bibinfo {author} {\bibfnamefont {R.}~\bibnamefont {Gastmans}},\
  }\bibfield  {title} {\bibinfo {title} {{Conformal Invariance in Quantum
  Gravity}},\ }\href {https://doi.org/10.1016/0550-3213(76)90406-5} {\bibfield
  {journal} {\bibinfo  {journal} {Nucl. Phys. B}\ }\textbf {\bibinfo {volume}
  {117}},\ \bibinfo {pages} {407} (\bibinfo {year} {1976})}\BibitemShut
  {NoStop}%
\bibitem [{\citenamefont {Shaposhnikov}\ and\ \citenamefont
  {Zenhausern}(2009{\natexlab{a}})}]{Shaposhnikov:2008xb}%
  \BibitemOpen
  \bibfield  {author} {\bibinfo {author} {\bibfnamefont {M.}~\bibnamefont
  {Shaposhnikov}}\ and\ \bibinfo {author} {\bibfnamefont {D.}~\bibnamefont
  {Zenhausern}},\ }\bibfield  {title} {\bibinfo {title} {{Scale invariance,
  unimodular gravity and dark energy}},\ }\href
  {https://doi.org/10.1016/j.physletb.2008.11.054} {\bibfield  {journal}
  {\bibinfo  {journal} {Phys. Lett. B}\ }\textbf {\bibinfo {volume} {671}},\
  \bibinfo {pages} {187} (\bibinfo {year} {2009}{\natexlab{a}})},\ \Eprint
  {https://arxiv.org/abs/0809.3395} {arXiv:0809.3395 [hep-th]} \BibitemShut
  {NoStop}%
\bibitem [{\citenamefont {Shaposhnikov}\ and\ \citenamefont
  {Zenhausern}(2009{\natexlab{b}})}]{Shaposhnikov:2008xi}%
  \BibitemOpen
  \bibfield  {author} {\bibinfo {author} {\bibfnamefont {M.}~\bibnamefont
  {Shaposhnikov}}\ and\ \bibinfo {author} {\bibfnamefont {D.}~\bibnamefont
  {Zenhausern}},\ }\bibfield  {title} {\bibinfo {title} {{Quantum scale
  invariance, cosmological constant and hierarchy problem}},\ }\href
  {https://doi.org/10.1016/j.physletb.2008.11.041} {\bibfield  {journal}
  {\bibinfo  {journal} {Phys. Lett. B}\ }\textbf {\bibinfo {volume} {671}},\
  \bibinfo {pages} {162} (\bibinfo {year} {2009}{\natexlab{b}})},\ \Eprint
  {https://arxiv.org/abs/0809.3406} {arXiv:0809.3406 [hep-th]} \BibitemShut
  {NoStop}%
\bibitem [{\citenamefont {Gretsch}\ and\ \citenamefont
  {Monin}(2015)}]{Gretsch:2013ooa}%
  \BibitemOpen
  \bibfield  {author} {\bibinfo {author} {\bibfnamefont {F.}~\bibnamefont
  {Gretsch}}\ and\ \bibinfo {author} {\bibfnamefont {A.}~\bibnamefont
  {Monin}},\ }\bibfield  {title} {\bibinfo {title} {{Perturbative conformal
  symmetry and dilaton}},\ }\href {https://doi.org/10.1103/PhysRevD.92.045036}
  {\bibfield  {journal} {\bibinfo  {journal} {Phys. Rev. D}\ }\textbf {\bibinfo
  {volume} {92}},\ \bibinfo {pages} {045036} (\bibinfo {year} {2015})},\
  \Eprint {https://arxiv.org/abs/1308.3863} {arXiv:1308.3863 [hep-th]}
  \BibitemShut {NoStop}%
\bibitem [{\citenamefont {Garcia-Bellido}\ \emph {et~al.}(2011)\citenamefont
  {Garcia-Bellido}, \citenamefont {Rubio}, \citenamefont {Shaposhnikov},\ and\
  \citenamefont {Zenhausern}}]{Garcia-Bellido:2011kqb}%
  \BibitemOpen
  \bibfield  {author} {\bibinfo {author} {\bibfnamefont {J.}~\bibnamefont
  {Garcia-Bellido}}, \bibinfo {author} {\bibfnamefont {J.}~\bibnamefont
  {Rubio}}, \bibinfo {author} {\bibfnamefont {M.}~\bibnamefont
  {Shaposhnikov}},\ and\ \bibinfo {author} {\bibfnamefont {D.}~\bibnamefont
  {Zenhausern}},\ }\bibfield  {title} {\bibinfo {title} {{Higgs-Dilaton
  Cosmology: From the Early to the Late Universe}},\ }\href
  {https://doi.org/10.1103/PhysRevD.84.123504} {\bibfield  {journal} {\bibinfo
  {journal} {Phys. Rev. D}\ }\textbf {\bibinfo {volume} {84}},\ \bibinfo
  {pages} {123504} (\bibinfo {year} {2011})},\ \Eprint
  {https://arxiv.org/abs/1107.2163} {arXiv:1107.2163 [hep-ph]} \BibitemShut
  {NoStop}%
\bibitem [{\citenamefont {Bezrukov}\ \emph {et~al.}(2013)\citenamefont
  {Bezrukov}, \citenamefont {Karananas}, \citenamefont {Rubio},\ and\
  \citenamefont {Shaposhnikov}}]{Bezrukov:2012hx}%
  \BibitemOpen
  \bibfield  {author} {\bibinfo {author} {\bibfnamefont {F.}~\bibnamefont
  {Bezrukov}}, \bibinfo {author} {\bibfnamefont {G.~K.}\ \bibnamefont
  {Karananas}}, \bibinfo {author} {\bibfnamefont {J.}~\bibnamefont {Rubio}},\
  and\ \bibinfo {author} {\bibfnamefont {M.}~\bibnamefont {Shaposhnikov}},\
  }\bibfield  {title} {\bibinfo {title} {{Higgs-Dilaton Cosmology: an effective
  field theory approach}},\ }\href {https://doi.org/10.1103/PhysRevD.87.096001}
  {\bibfield  {journal} {\bibinfo  {journal} {Phys. Rev. D}\ }\textbf {\bibinfo
  {volume} {87}},\ \bibinfo {pages} {096001} (\bibinfo {year} {2013})},\
  \Eprint {https://arxiv.org/abs/1212.4148} {arXiv:1212.4148 [hep-ph]}
  \BibitemShut {NoStop}%
\bibitem [{\citenamefont {Shaposhnikov}\ and\ \citenamefont
  {Tokareva}(2022{\natexlab{a}})}]{Shaposhnikov:2022dou}%
  \BibitemOpen
  \bibfield  {author} {\bibinfo {author} {\bibfnamefont {M.}~\bibnamefont
  {Shaposhnikov}}\ and\ \bibinfo {author} {\bibfnamefont {A.}~\bibnamefont
  {Tokareva}},\ }\bibfield  {title} {\bibinfo {title} {{Anomaly-free scale
  symmetry and gravity}},\ }\href@noop {} {\  (\bibinfo {year}
  {2022}{\natexlab{a}})},\ \Eprint {https://arxiv.org/abs/2201.09232}
  {arXiv:2201.09232 [hep-th]} \BibitemShut {NoStop}%
\bibitem [{\citenamefont {Shaposhnikov}\ and\ \citenamefont
  {Tokareva}(2022{\natexlab{b}})}]{inpreparation}%
  \BibitemOpen
  \bibfield  {author} {\bibinfo {author} {\bibfnamefont {M.}~\bibnamefont
  {Shaposhnikov}}\ and\ \bibinfo {author} {\bibfnamefont {A.}~\bibnamefont
  {Tokareva}},\ }\bibfield  {title} {\bibinfo {title} {{Exact quantum conformal
  symmetry, its spontaneous breakdown and gravitational Weyl anomaly}},\
  }\href@noop {} {\bibfield  {journal} {\bibinfo  {journal} {In preparation}\ }
  (\bibinfo {year} {2022}{\natexlab{b}})}\BibitemShut {NoStop}%
\bibitem [{\citenamefont {Shaposhnikov}\ and\ \citenamefont
  {Tkachov}(2009)}]{Shaposhnikov:2009nk}%
  \BibitemOpen
  \bibfield  {author} {\bibinfo {author} {\bibfnamefont {M.~E.}\ \bibnamefont
  {Shaposhnikov}}\ and\ \bibinfo {author} {\bibfnamefont {F.~V.}\ \bibnamefont
  {Tkachov}},\ }\bibfield  {title} {\bibinfo {title} {{Quantum scale-invariant
  models as effective field theories}},\ }\href@noop {} {\  (\bibinfo {year}
  {2009})},\ \Eprint {https://arxiv.org/abs/0905.4857} {arXiv:0905.4857
  [hep-th]} \BibitemShut {NoStop}%
\bibitem [{\citenamefont {Karananas}\ and\ \citenamefont
  {Shaposhnikov}(2018)}]{Karananas:2017zrg}%
  \BibitemOpen
  \bibfield  {author} {\bibinfo {author} {\bibfnamefont {G.~K.}\ \bibnamefont
  {Karananas}}\ and\ \bibinfo {author} {\bibfnamefont {M.}~\bibnamefont
  {Shaposhnikov}},\ }\bibfield  {title} {\bibinfo {title} {{CFT data and
  spontaneously broken conformal invariance}},\ }\href
  {https://doi.org/10.1103/PhysRevD.97.045009} {\bibfield  {journal} {\bibinfo
  {journal} {Phys. Rev. D}\ }\textbf {\bibinfo {volume} {97}},\ \bibinfo
  {pages} {045009} (\bibinfo {year} {2018})},\ \Eprint
  {https://arxiv.org/abs/1708.02220} {arXiv:1708.02220 [hep-th]} \BibitemShut
  {NoStop}%
\bibitem [{\citenamefont {Garcia-Bellido}\ \emph {et~al.}(2012)\citenamefont
  {Garcia-Bellido}, \citenamefont {Rubio},\ and\ \citenamefont
  {Shaposhnikov}}]{Garcia-Bellido:2012npk}%
  \BibitemOpen
  \bibfield  {author} {\bibinfo {author} {\bibfnamefont {J.}~\bibnamefont
  {Garcia-Bellido}}, \bibinfo {author} {\bibfnamefont {J.}~\bibnamefont
  {Rubio}},\ and\ \bibinfo {author} {\bibfnamefont {M.}~\bibnamefont
  {Shaposhnikov}},\ }\bibfield  {title} {\bibinfo {title} {{Higgs-Dilaton
  cosmology: Are there extra relativistic species?}},\ }\href
  {https://doi.org/10.1016/j.physletb.2012.10.075} {\bibfield  {journal}
  {\bibinfo  {journal} {Phys. Lett. B}\ }\textbf {\bibinfo {volume} {718}},\
  \bibinfo {pages} {507} (\bibinfo {year} {2012})},\ \Eprint
  {https://arxiv.org/abs/1209.2119} {arXiv:1209.2119 [hep-ph]} \BibitemShut
  {NoStop}%
\bibitem [{\citenamefont {Wetterich}(1988{\natexlab{a}})}]{Wetterich:1987fm}%
  \BibitemOpen
  \bibfield  {author} {\bibinfo {author} {\bibfnamefont {C.}~\bibnamefont
  {Wetterich}},\ }\bibfield  {title} {\bibinfo {title} {{Cosmology and the Fate
  of Dilatation Symmetry}},\ }\href
  {https://doi.org/10.1016/0550-3213(88)90193-9} {\bibfield  {journal}
  {\bibinfo  {journal} {Nucl. Phys. B}\ }\textbf {\bibinfo {volume} {302}},\
  \bibinfo {pages} {668} (\bibinfo {year} {1988}{\natexlab{a}})},\ \Eprint
  {https://arxiv.org/abs/1711.03844} {arXiv:1711.03844 [hep-th]} \BibitemShut
  {NoStop}%
\bibitem [{\citenamefont {Wetterich}(1988{\natexlab{b}})}]{Wetterich:1987fk}%
  \BibitemOpen
  \bibfield  {author} {\bibinfo {author} {\bibfnamefont {C.}~\bibnamefont
  {Wetterich}},\ }\bibfield  {title} {\bibinfo {title} {{Cosmologies With
  Variable Newton's 'Constant'}},\ }\href
  {https://doi.org/10.1016/0550-3213(88)90192-7} {\bibfield  {journal}
  {\bibinfo  {journal} {Nucl. Phys. B}\ }\textbf {\bibinfo {volume} {302}},\
  \bibinfo {pages} {645} (\bibinfo {year} {1988}{\natexlab{b}})}\BibitemShut
  {NoStop}%
\bibitem [{\citenamefont {Ferreira}\ \emph {et~al.}(2017)\citenamefont
  {Ferreira}, \citenamefont {Hill},\ and\ \citenamefont
  {Ross}}]{Ferreira:2016kxi}%
  \BibitemOpen
  \bibfield  {author} {\bibinfo {author} {\bibfnamefont {P.~G.}\ \bibnamefont
  {Ferreira}}, \bibinfo {author} {\bibfnamefont {C.~T.}\ \bibnamefont {Hill}},\
  and\ \bibinfo {author} {\bibfnamefont {G.~G.}\ \bibnamefont {Ross}},\
  }\bibfield  {title} {\bibinfo {title} {{No fifth force in a scale invariant
  universe}},\ }\href {https://doi.org/10.1103/PhysRevD.95.064038} {\bibfield
  {journal} {\bibinfo  {journal} {Phys. Rev. D}\ }\textbf {\bibinfo {volume}
  {95}},\ \bibinfo {pages} {064038} (\bibinfo {year} {2017})},\ \Eprint
  {https://arxiv.org/abs/1612.03157} {arXiv:1612.03157 [gr-qc]} \BibitemShut
  {NoStop}%
\bibitem [{\citenamefont {Marsh}(2016)}]{Marsh:2015xka}%
  \BibitemOpen
  \bibfield  {author} {\bibinfo {author} {\bibfnamefont {D.~J.~E.}\
  \bibnamefont {Marsh}},\ }\bibfield  {title} {\bibinfo {title} {{Axion
  Cosmology}},\ }\href {https://doi.org/10.1016/j.physrep.2016.06.005}
  {\bibfield  {journal} {\bibinfo  {journal} {Phys. Rept.}\ }\textbf {\bibinfo
  {volume} {643}},\ \bibinfo {pages} {1} (\bibinfo {year} {2016})},\ \Eprint
  {https://arxiv.org/abs/1510.07633} {arXiv:1510.07633 [astro-ph.CO]}
  \BibitemShut {NoStop}%
\bibitem [{\citenamefont {Grilli~di Cortona}\ \emph {et~al.}(2016)\citenamefont
  {Grilli~di Cortona}, \citenamefont {Hardy}, \citenamefont {Pardo~Vega},\ and\
  \citenamefont {Villadoro}}]{GrillidiCortona:2015jxo}%
  \BibitemOpen
  \bibfield  {author} {\bibinfo {author} {\bibfnamefont {G.}~\bibnamefont
  {Grilli~di Cortona}}, \bibinfo {author} {\bibfnamefont {E.}~\bibnamefont
  {Hardy}}, \bibinfo {author} {\bibfnamefont {J.}~\bibnamefont {Pardo~Vega}},\
  and\ \bibinfo {author} {\bibfnamefont {G.}~\bibnamefont {Villadoro}},\
  }\bibfield  {title} {\bibinfo {title} {{The QCD axion, precisely}},\ }\href
  {https://doi.org/10.1007/JHEP01(2016)034} {\bibfield  {journal} {\bibinfo
  {journal} {JHEP}\ }\textbf {\bibinfo {volume} {01}},\ \bibinfo {pages}
  {034}},\ \Eprint {https://arxiv.org/abs/1511.02867} {arXiv:1511.02867
  [hep-ph]} \BibitemShut {NoStop}%
\bibitem [{\citenamefont
  {Schmidt-Wellenburg}(2016)}]{Schmidt-Wellenburg:2016nfv}%
  \BibitemOpen
  \bibfield  {author} {\bibinfo {author} {\bibfnamefont {P.}~\bibnamefont
  {Schmidt-Wellenburg}},\ }\bibfield  {title} {\bibinfo {title} {{The quest to
  find an electric dipole moment of the neutron}},\ }\href@noop {} {\
  (\bibinfo {year} {2016})},\ \Eprint {https://arxiv.org/abs/1607.06609}
  {arXiv:1607.06609 [hep-ex]} \BibitemShut {NoStop}%
\bibitem [{\citenamefont {Pendlebury}\ \emph {et~al.}(2015)\citenamefont
  {Pendlebury} \emph {et~al.}}]{Pendlebury:2015lrz}%
  \BibitemOpen
  \bibfield  {author} {\bibinfo {author} {\bibfnamefont {J.~M.}\ \bibnamefont
  {Pendlebury}} \emph {et~al.},\ }\bibfield  {title} {\bibinfo {title}
  {{Revised experimental upper limit on the electric dipole moment of the
  neutron}},\ }\href {https://doi.org/10.1103/PhysRevD.92.092003} {\bibfield
  {journal} {\bibinfo  {journal} {Phys. Rev. D}\ }\textbf {\bibinfo {volume}
  {92}},\ \bibinfo {pages} {092003} (\bibinfo {year} {2015})},\ \Eprint
  {https://arxiv.org/abs/1509.04411} {arXiv:1509.04411 [hep-ex]} \BibitemShut
  {NoStop}%
\bibitem [{\citenamefont {Peccei}\ and\ \citenamefont
  {Quinn}(1977)}]{Peccei:1977hh}%
  \BibitemOpen
  \bibfield  {author} {\bibinfo {author} {\bibfnamefont {R.~D.}\ \bibnamefont
  {Peccei}}\ and\ \bibinfo {author} {\bibfnamefont {H.~R.}\ \bibnamefont
  {Quinn}},\ }\bibfield  {title} {\bibinfo {title} {{CP Conservation in the
  Presence of Instantons}},\ }\href
  {https://doi.org/10.1103/PhysRevLett.38.1440} {\bibfield  {journal} {\bibinfo
   {journal} {Phys. Rev. Lett.}\ }\textbf {\bibinfo {volume} {38}},\ \bibinfo
  {pages} {1440} (\bibinfo {year} {1977})}\BibitemShut {NoStop}%
\bibitem [{\citenamefont {Di~Vecchia}\ and\ \citenamefont
  {Veneziano}(1980)}]{DiVecchia:1980yfw}%
  \BibitemOpen
  \bibfield  {author} {\bibinfo {author} {\bibfnamefont {P.}~\bibnamefont
  {Di~Vecchia}}\ and\ \bibinfo {author} {\bibfnamefont {G.}~\bibnamefont
  {Veneziano}},\ }\bibfield  {title} {\bibinfo {title} {{Chiral Dynamics in the
  Large n Limit}},\ }\href {https://doi.org/10.1016/0550-3213(80)90370-3}
  {\bibfield  {journal} {\bibinfo  {journal} {Nucl. Phys. B}\ }\textbf
  {\bibinfo {volume} {171}},\ \bibinfo {pages} {253} (\bibinfo {year}
  {1980})}\BibitemShut {NoStop}%
\bibitem [{\citenamefont {Kim}(1979)}]{Kim:1979if}%
  \BibitemOpen
  \bibfield  {author} {\bibinfo {author} {\bibfnamefont {J.~E.}\ \bibnamefont
  {Kim}},\ }\bibfield  {title} {\bibinfo {title} {{Weak Interaction Singlet and
  Strong CP Invariance}},\ }\href {https://doi.org/10.1103/PhysRevLett.43.103}
  {\bibfield  {journal} {\bibinfo  {journal} {Phys. Rev. Lett.}\ }\textbf
  {\bibinfo {volume} {43}},\ \bibinfo {pages} {103} (\bibinfo {year}
  {1979})}\BibitemShut {NoStop}%
\bibitem [{\citenamefont {Shifman}\ \emph {et~al.}(1980)\citenamefont
  {Shifman}, \citenamefont {Vainshtein},\ and\ \citenamefont
  {Zakharov}}]{Shifman:1979if}%
  \BibitemOpen
  \bibfield  {author} {\bibinfo {author} {\bibfnamefont {M.~A.}\ \bibnamefont
  {Shifman}}, \bibinfo {author} {\bibfnamefont {A.~I.}\ \bibnamefont
  {Vainshtein}},\ and\ \bibinfo {author} {\bibfnamefont {V.~I.}\ \bibnamefont
  {Zakharov}},\ }\bibfield  {title} {\bibinfo {title} {{Can Confinement Ensure
  Natural CP Invariance of Strong Interactions?}},\ }\href
  {https://doi.org/10.1016/0550-3213(80)90209-6} {\bibfield  {journal}
  {\bibinfo  {journal} {Nucl. Phys. B}\ }\textbf {\bibinfo {volume} {166}},\
  \bibinfo {pages} {493} (\bibinfo {year} {1980})}\BibitemShut {NoStop}%
\bibitem [{\citenamefont {Dine}\ \emph {et~al.}(1981)\citenamefont {Dine},
  \citenamefont {Fischler},\ and\ \citenamefont {Srednicki}}]{Dine:1981rt}%
  \BibitemOpen
  \bibfield  {author} {\bibinfo {author} {\bibfnamefont {M.}~\bibnamefont
  {Dine}}, \bibinfo {author} {\bibfnamefont {W.}~\bibnamefont {Fischler}},\
  and\ \bibinfo {author} {\bibfnamefont {M.}~\bibnamefont {Srednicki}},\
  }\bibfield  {title} {\bibinfo {title} {{A Simple Solution to the Strong CP
  Problem with a Harmless Axion}},\ }\href
  {https://doi.org/10.1016/0370-2693(81)90590-6} {\bibfield  {journal}
  {\bibinfo  {journal} {Phys. Lett. B}\ }\textbf {\bibinfo {volume} {104}},\
  \bibinfo {pages} {199} (\bibinfo {year} {1981})}\BibitemShut {NoStop}%
\bibitem [{\citenamefont {Abbott}\ and\ \citenamefont
  {Sikivie}(1983)}]{Abbott:1982af}%
  \BibitemOpen
  \bibfield  {author} {\bibinfo {author} {\bibfnamefont {L.~F.}\ \bibnamefont
  {Abbott}}\ and\ \bibinfo {author} {\bibfnamefont {P.}~\bibnamefont
  {Sikivie}},\ }\bibfield  {title} {\bibinfo {title} {{A Cosmological Bound on
  the Invisible Axion}},\ }\href {https://doi.org/10.1016/0370-2693(83)90638-X}
  {\bibfield  {journal} {\bibinfo  {journal} {Phys. Lett. B}\ }\textbf
  {\bibinfo {volume} {120}},\ \bibinfo {pages} {133} (\bibinfo {year}
  {1983})}\BibitemShut {NoStop}%
\bibitem [{\citenamefont {Preskill}\ \emph {et~al.}(1983)\citenamefont
  {Preskill}, \citenamefont {Wise},\ and\ \citenamefont
  {Wilczek}}]{Preskill:1982cy}%
  \BibitemOpen
  \bibfield  {author} {\bibinfo {author} {\bibfnamefont {J.}~\bibnamefont
  {Preskill}}, \bibinfo {author} {\bibfnamefont {M.~B.}\ \bibnamefont {Wise}},\
  and\ \bibinfo {author} {\bibfnamefont {F.}~\bibnamefont {Wilczek}},\
  }\bibfield  {title} {\bibinfo {title} {{Cosmology of the Invisible Axion}},\
  }\href {https://doi.org/10.1016/0370-2693(83)90637-8} {\bibfield  {journal}
  {\bibinfo  {journal} {Phys. Lett. B}\ }\textbf {\bibinfo {volume} {120}},\
  \bibinfo {pages} {127} (\bibinfo {year} {1983})}\BibitemShut {NoStop}%
\bibitem [{\citenamefont {Lyth}(1990)}]{Lyth:1989pb}%
  \BibitemOpen
  \bibfield  {author} {\bibinfo {author} {\bibfnamefont {D.~H.}\ \bibnamefont
  {Lyth}},\ }\bibfield  {title} {\bibinfo {title} {{A Limit on the Inflationary
  Energy Density From Axion Isocurvature Fluctuations}},\ }\href
  {https://doi.org/10.1016/0370-2693(90)90374-F} {\bibfield  {journal}
  {\bibinfo  {journal} {Phys. Lett. B}\ }\textbf {\bibinfo {volume} {236}},\
  \bibinfo {pages} {408} (\bibinfo {year} {1990})}\BibitemShut {NoStop}%
\bibitem [{\citenamefont {Akrami}\ \emph {et~al.}(2020)\citenamefont {Akrami}
  \emph {et~al.}}]{Planck:2018jri}%
  \BibitemOpen
  \bibfield  {author} {\bibinfo {author} {\bibfnamefont {Y.}~\bibnamefont
  {Akrami}} \emph {et~al.} (\bibinfo {collaboration} {Planck}),\ }\bibfield
  {title} {\bibinfo {title} {{Planck 2018 results. X. Constraints on
  inflation}},\ }\href {https://doi.org/10.1051/0004-6361/201833887} {\bibfield
   {journal} {\bibinfo  {journal} {Astron. Astrophys.}\ }\textbf {\bibinfo
  {volume} {641}},\ \bibinfo {pages} {A10} (\bibinfo {year} {2020})},\ \Eprint
  {https://arxiv.org/abs/1807.06211} {arXiv:1807.06211 [astro-ph.CO]}
  \BibitemShut {NoStop}%
\bibitem [{\citenamefont {Aghanim}\ \emph {et~al.}(2020)\citenamefont {Aghanim}
  \emph {et~al.}}]{Planck:2018vyg}%
  \BibitemOpen
  \bibfield  {author} {\bibinfo {author} {\bibfnamefont {N.}~\bibnamefont
  {Aghanim}} \emph {et~al.} (\bibinfo {collaboration} {Planck}),\ }\bibfield
  {title} {\bibinfo {title} {{Planck 2018 results. VI. Cosmological
  parameters}},\ }\href {https://doi.org/10.1051/0004-6361/201833910}
  {\bibfield  {journal} {\bibinfo  {journal} {Astron. Astrophys.}\ }\textbf
  {\bibinfo {volume} {641}},\ \bibinfo {pages} {A6} (\bibinfo {year} {2020})},\
  \bibinfo {note} {[Erratum: Astron.Astrophys. 652, C4 (2021)]},\ \Eprint
  {https://arxiv.org/abs/1807.06209} {arXiv:1807.06209 [astro-ph.CO]}
  \BibitemShut {NoStop}%
\bibitem [{\citenamefont {Takahashi}\ \emph {et~al.}(2018)\citenamefont
  {Takahashi}, \citenamefont {Yin},\ and\ \citenamefont
  {Guth}}]{Takahashi:2018tdu}%
  \BibitemOpen
  \bibfield  {author} {\bibinfo {author} {\bibfnamefont {F.}~\bibnamefont
  {Takahashi}}, \bibinfo {author} {\bibfnamefont {W.}~\bibnamefont {Yin}},\
  and\ \bibinfo {author} {\bibfnamefont {A.~H.}\ \bibnamefont {Guth}},\
  }\bibfield  {title} {\bibinfo {title} {{QCD axion window and low-scale
  inflation}},\ }\href {https://doi.org/10.1103/PhysRevD.98.015042} {\bibfield
  {journal} {\bibinfo  {journal} {Phys. Rev. D}\ }\textbf {\bibinfo {volume}
  {98}},\ \bibinfo {pages} {015042} (\bibinfo {year} {2018})},\ \Eprint
  {https://arxiv.org/abs/1805.08763} {arXiv:1805.08763 [hep-ph]} \BibitemShut
  {NoStop}%
\bibitem [{\citenamefont {Kearney}\ \emph {et~al.}(2016)\citenamefont
  {Kearney}, \citenamefont {Orlofsky},\ and\ \citenamefont
  {Pierce}}]{Kearney:2016vqw}%
  \BibitemOpen
  \bibfield  {author} {\bibinfo {author} {\bibfnamefont {J.}~\bibnamefont
  {Kearney}}, \bibinfo {author} {\bibfnamefont {N.}~\bibnamefont {Orlofsky}},\
  and\ \bibinfo {author} {\bibfnamefont {A.}~\bibnamefont {Pierce}},\
  }\bibfield  {title} {\bibinfo {title} {{High-Scale Axions without
  Isocurvature from Inflationary Dynamics}},\ }\href
  {https://doi.org/10.1103/PhysRevD.93.095026} {\bibfield  {journal} {\bibinfo
  {journal} {Phys. Rev. D}\ }\textbf {\bibinfo {volume} {93}},\ \bibinfo
  {pages} {095026} (\bibinfo {year} {2016})},\ \Eprint
  {https://arxiv.org/abs/1601.03049} {arXiv:1601.03049 [hep-ph]} \BibitemShut
  {NoStop}%
\bibitem [{\citenamefont {Ballesteros}\ \emph {et~al.}(2021)\citenamefont
  {Ballesteros}, \citenamefont {Ringwald}, \citenamefont {Tamarit},\ and\
  \citenamefont {Welling}}]{Ballesteros:2021bee}%
  \BibitemOpen
  \bibfield  {author} {\bibinfo {author} {\bibfnamefont {G.}~\bibnamefont
  {Ballesteros}}, \bibinfo {author} {\bibfnamefont {A.}~\bibnamefont
  {Ringwald}}, \bibinfo {author} {\bibfnamefont {C.}~\bibnamefont {Tamarit}},\
  and\ \bibinfo {author} {\bibfnamefont {Y.}~\bibnamefont {Welling}},\
  }\bibfield  {title} {\bibinfo {title} {{Revisiting isocurvature bounds in
  models unifying the axion with the inflaton}},\ }\href
  {https://doi.org/10.1088/1475-7516/2021/09/036} {\bibfield  {journal}
  {\bibinfo  {journal} {JCAP}\ }\textbf {\bibinfo {volume} {09}},\ \bibinfo
  {pages} {036}},\ \Eprint {https://arxiv.org/abs/2104.13847} {arXiv:2104.13847
  [hep-ph]} \BibitemShut {NoStop}%
\bibitem [{\citenamefont {Almeida}\ \emph {et~al.}(2019)\citenamefont
  {Almeida}, \citenamefont {Bernal}, \citenamefont {Rubio},\ and\ \citenamefont
  {Tenkanen}}]{Almeida:2018oid}%
  \BibitemOpen
  \bibfield  {author} {\bibinfo {author} {\bibfnamefont {J.~P.~B.}\
  \bibnamefont {Almeida}}, \bibinfo {author} {\bibfnamefont {N.}~\bibnamefont
  {Bernal}}, \bibinfo {author} {\bibfnamefont {J.}~\bibnamefont {Rubio}},\ and\
  \bibinfo {author} {\bibfnamefont {T.}~\bibnamefont {Tenkanen}},\ }\bibfield
  {title} {\bibinfo {title} {{Hidden inflation dark matter}},\ }\href
  {https://doi.org/10.1088/1475-7516/2019/03/012} {\bibfield  {journal}
  {\bibinfo  {journal} {JCAP}\ }\textbf {\bibinfo {volume} {03}},\ \bibinfo
  {pages} {012}},\ \Eprint {https://arxiv.org/abs/1811.09640} {arXiv:1811.09640
  [hep-ph]} \BibitemShut {NoStop}%
\bibitem [{\citenamefont {Rubio}\ and\ \citenamefont
  {Tomberg}(2019)}]{Rubio:2019ypq}%
  \BibitemOpen
  \bibfield  {author} {\bibinfo {author} {\bibfnamefont {J.}~\bibnamefont
  {Rubio}}\ and\ \bibinfo {author} {\bibfnamefont {E.~S.}\ \bibnamefont
  {Tomberg}},\ }\bibfield  {title} {\bibinfo {title} {{Preheating in Palatini
  Higgs inflation}},\ }\href {https://doi.org/10.1088/1475-7516/2019/04/021}
  {\bibfield  {journal} {\bibinfo  {journal} {JCAP}\ }\textbf {\bibinfo
  {volume} {04}},\ \bibinfo {pages} {021}},\ \Eprint
  {https://arxiv.org/abs/1902.10148} {arXiv:1902.10148 [hep-ph]} \BibitemShut
  {NoStop}%
\bibitem [{\citenamefont {Shaposhnikov}\ \emph {et~al.}(2020)\citenamefont
  {Shaposhnikov}, \citenamefont {Shkerin},\ and\ \citenamefont
  {Zell}}]{Shaposhnikov:2020fdv}%
  \BibitemOpen
  \bibfield  {author} {\bibinfo {author} {\bibfnamefont {M.}~\bibnamefont
  {Shaposhnikov}}, \bibinfo {author} {\bibfnamefont {A.}~\bibnamefont
  {Shkerin}},\ and\ \bibinfo {author} {\bibfnamefont {S.}~\bibnamefont
  {Zell}},\ }\bibfield  {title} {\bibinfo {title} {{Quantum Effects in Palatini
  Higgs Inflation}},\ }\href {https://doi.org/10.1088/1475-7516/2020/07/064}
  {\bibfield  {journal} {\bibinfo  {journal} {JCAP}\ }\textbf {\bibinfo
  {volume} {07}},\ \bibinfo {pages} {064}},\ \Eprint
  {https://arxiv.org/abs/2002.07105} {arXiv:2002.07105 [hep-ph]} \BibitemShut
  {NoStop}%
\bibitem [{\citenamefont {Rasanen}\ and\ \citenamefont
  {Verbin}(2022)}]{Rasanen:2022ijc}%
  \BibitemOpen
  \bibfield  {author} {\bibinfo {author} {\bibfnamefont {S.}~\bibnamefont
  {Rasanen}}\ and\ \bibinfo {author} {\bibfnamefont {Y.}~\bibnamefont
  {Verbin}},\ }\bibfield  {title} {\bibinfo {title} {{Palatini formulation for
  gauge theory: implications for slow-roll inflation}},\ }\href@noop {} {\
  (\bibinfo {year} {2022})},\ \Eprint {https://arxiv.org/abs/2211.15584}
  {arXiv:2211.15584 [astro-ph.CO]} \BibitemShut {NoStop}%
\bibitem [{\citenamefont {Ito}\ \emph {et~al.}(2022)\citenamefont {Ito},
  \citenamefont {Khater},\ and\ \citenamefont {Rasanen}}]{Ito:2021ssc}%
  \BibitemOpen
  \bibfield  {author} {\bibinfo {author} {\bibfnamefont {A.}~\bibnamefont
  {Ito}}, \bibinfo {author} {\bibfnamefont {W.}~\bibnamefont {Khater}},\ and\
  \bibinfo {author} {\bibfnamefont {S.}~\bibnamefont {Rasanen}},\ }\bibfield
  {title} {\bibinfo {title} {{Tree-level unitarity in Higgs inflation in the
  metric and the Palatini formulation}},\ }\href
  {https://doi.org/10.1007/JHEP06(2022)164} {\bibfield  {journal} {\bibinfo
  {journal} {JHEP}\ }\textbf {\bibinfo {volume} {06}},\ \bibinfo {pages}
  {164}},\ \Eprint {https://arxiv.org/abs/2111.05621} {arXiv:2111.05621
  [astro-ph.CO]} \BibitemShut {NoStop}%
\bibitem [{\citenamefont {Rasanen}(2019)}]{Rasanen:2018ihz}%
  \BibitemOpen
  \bibfield  {author} {\bibinfo {author} {\bibfnamefont {S.}~\bibnamefont
  {Rasanen}},\ }\bibfield  {title} {\bibinfo {title} {{Higgs inflation in the
  Palatini formulation with kinetic terms for the metric}},\ }\href
  {https://doi.org/10.21105/astro.1811.09514} {\bibfield  {journal} {\bibinfo
  {journal} {Open J. Astrophys.}\ }\textbf {\bibinfo {volume} {2}},\ \bibinfo
  {pages} {1} (\bibinfo {year} {2019})},\ \Eprint
  {https://arxiv.org/abs/1811.09514} {arXiv:1811.09514 [gr-qc]} \BibitemShut
  {NoStop}%
\bibitem [{\citenamefont {Enckell}\ \emph {et~al.}(2021)\citenamefont
  {Enckell}, \citenamefont {Nurmi}, \citenamefont {R\"as\"anen},\ and\
  \citenamefont {Tomberg}}]{Enckell:2020lvn}%
  \BibitemOpen
  \bibfield  {author} {\bibinfo {author} {\bibfnamefont {V.-M.}\ \bibnamefont
  {Enckell}}, \bibinfo {author} {\bibfnamefont {S.}~\bibnamefont {Nurmi}},
  \bibinfo {author} {\bibfnamefont {S.}~\bibnamefont {R\"as\"anen}},\ and\
  \bibinfo {author} {\bibfnamefont {E.}~\bibnamefont {Tomberg}},\ }\bibfield
  {title} {\bibinfo {title} {{Critical point Higgs inflation in the Palatini
  formulation}},\ }\href {https://doi.org/10.1007/JHEP04(2021)059} {\bibfield
  {journal} {\bibinfo  {journal} {JHEP}\ }\textbf {\bibinfo {volume} {04}},\
  \bibinfo {pages} {059}},\ \Eprint {https://arxiv.org/abs/2012.03660}
  {arXiv:2012.03660 [astro-ph.CO]} \BibitemShut {NoStop}%
\bibitem [{\citenamefont {Piani}\ and\ \citenamefont
  {Rubio}(2022)}]{Piani:2022gon}%
  \BibitemOpen
  \bibfield  {author} {\bibinfo {author} {\bibfnamefont {M.}~\bibnamefont
  {Piani}}\ and\ \bibinfo {author} {\bibfnamefont {J.}~\bibnamefont {Rubio}},\
  }\bibfield  {title} {\bibinfo {title} {{Higgs-Dilaton inflation in
  Einstein-Cartan gravity}},\ }\href
  {https://doi.org/10.1088/1475-7516/2022/05/009} {\bibfield  {journal}
  {\bibinfo  {journal} {JCAP}\ }\textbf {\bibinfo {volume} {05}}\bibfield
  {number} {\bibinfo  {number} { (05)},\ \bibinfo {pages} {009}},\ }\Eprint
  {https://arxiv.org/abs/2202.04665} {arXiv:2202.04665 [gr-qc]} \BibitemShut
  {NoStop}%
\bibitem [{\citenamefont {Yin}(2022)}]{Yin:2022fgo}%
  \BibitemOpen
  \bibfield  {author} {\bibinfo {author} {\bibfnamefont {W.}~\bibnamefont
  {Yin}},\ }\bibfield  {title} {\bibinfo {title} {{Weak-Scale Higgs
  Inflation}},\ }\href@noop {} {\  (\bibinfo {year} {2022})},\ \Eprint
  {https://arxiv.org/abs/2210.15680} {arXiv:2210.15680 [hep-ph]} \BibitemShut
  {NoStop}%
\bibitem [{\citenamefont {Bezrukov}\ \emph {et~al.}(2011)\citenamefont
  {Bezrukov}, \citenamefont {Magnin}, \citenamefont {Shaposhnikov},\ and\
  \citenamefont {Sibiryakov}}]{Bezrukov:2010jz}%
  \BibitemOpen
  \bibfield  {author} {\bibinfo {author} {\bibfnamefont {F.}~\bibnamefont
  {Bezrukov}}, \bibinfo {author} {\bibfnamefont {A.}~\bibnamefont {Magnin}},
  \bibinfo {author} {\bibfnamefont {M.}~\bibnamefont {Shaposhnikov}},\ and\
  \bibinfo {author} {\bibfnamefont {S.}~\bibnamefont {Sibiryakov}},\ }\bibfield
   {title} {\bibinfo {title} {{Higgs inflation: consistency and
  generalisations}},\ }\href {https://doi.org/10.1007/JHEP01(2011)016}
  {\bibfield  {journal} {\bibinfo  {journal} {JHEP}\ }\textbf {\bibinfo
  {volume} {01}},\ \bibinfo {pages} {016}},\ \Eprint
  {https://arxiv.org/abs/1008.5157} {arXiv:1008.5157 [hep-ph]} \BibitemShut
  {NoStop}%
\bibitem [{\citenamefont {Visinelli}\ and\ \citenamefont
  {Gondolo}(2014)}]{Visinelli:2014twa}%
  \BibitemOpen
  \bibfield  {author} {\bibinfo {author} {\bibfnamefont {L.}~\bibnamefont
  {Visinelli}}\ and\ \bibinfo {author} {\bibfnamefont {P.}~\bibnamefont
  {Gondolo}},\ }\bibfield  {title} {\bibinfo {title} {{Axion cold dark matter
  in view of BICEP2 results}},\ }\href
  {https://doi.org/10.1103/PhysRevLett.113.011802} {\bibfield  {journal}
  {\bibinfo  {journal} {Phys. Rev. Lett.}\ }\textbf {\bibinfo {volume} {113}},\
  \bibinfo {pages} {011802} (\bibinfo {year} {2014})},\ \Eprint
  {https://arxiv.org/abs/1403.4594} {arXiv:1403.4594 [hep-ph]} \BibitemShut
  {NoStop}%
\bibitem [{\citenamefont {Turner}(1986)}]{Turner:1985si}%
  \BibitemOpen
  \bibfield  {author} {\bibinfo {author} {\bibfnamefont {M.~S.}\ \bibnamefont
  {Turner}},\ }\bibfield  {title} {\bibinfo {title} {{Cosmic and Local Mass
  Density of Invisible Axions}},\ }\href
  {https://doi.org/10.1103/PhysRevD.33.889} {\bibfield  {journal} {\bibinfo
  {journal} {Phys. Rev. D}\ }\textbf {\bibinfo {volume} {33}},\ \bibinfo
  {pages} {889} (\bibinfo {year} {1986})}\BibitemShut {NoStop}%
\bibitem [{\citenamefont {Linde}(1991)}]{Linde:1991km}%
  \BibitemOpen
  \bibfield  {author} {\bibinfo {author} {\bibfnamefont {A.~D.}\ \bibnamefont
  {Linde}},\ }\bibfield  {title} {\bibinfo {title} {{Axions in inflationary
  cosmology}},\ }\href {https://doi.org/10.1016/0370-2693(91)90130-I}
  {\bibfield  {journal} {\bibinfo  {journal} {Phys. Lett. B}\ }\textbf
  {\bibinfo {volume} {259}},\ \bibinfo {pages} {38} (\bibinfo {year}
  {1991})}\BibitemShut {NoStop}%
\bibitem [{\citenamefont {Linde}\ and\ \citenamefont
  {Lyth}(1990)}]{Linde:1990yj}%
  \BibitemOpen
  \bibfield  {author} {\bibinfo {author} {\bibfnamefont {A.~D.}\ \bibnamefont
  {Linde}}\ and\ \bibinfo {author} {\bibfnamefont {D.~H.}\ \bibnamefont
  {Lyth}},\ }\bibfield  {title} {\bibinfo {title} {{Axionic domain wall
  production during inflation}},\ }\href
  {https://doi.org/10.1016/0370-2693(90)90613-B} {\bibfield  {journal}
  {\bibinfo  {journal} {Phys. Lett. B}\ }\textbf {\bibinfo {volume} {246}},\
  \bibinfo {pages} {353} (\bibinfo {year} {1990})}\BibitemShut {NoStop}%
\bibitem [{\citenamefont {Rubio}(2020)}]{Rubio:2020zht}%
  \BibitemOpen
  \bibfield  {author} {\bibinfo {author} {\bibfnamefont {J.}~\bibnamefont
  {Rubio}},\ }\bibfield  {title} {\bibinfo {title} {{Scale symmetry, the Higgs
  and the Cosmos}},\ }\href {https://doi.org/10.22323/1.376.0074} {\bibfield
  {journal} {\bibinfo  {journal} {PoS}\ }\textbf {\bibinfo {volume}
  {CORFU2019}},\ \bibinfo {pages} {074} (\bibinfo {year} {2020})},\ \Eprint
  {https://arxiv.org/abs/2004.00039} {arXiv:2004.00039 [gr-qc]} \BibitemShut
  {NoStop}%
\bibitem [{\citenamefont {Casas}\ \emph {et~al.}(2018)\citenamefont {Casas},
  \citenamefont {Pauly},\ and\ \citenamefont {Rubio}}]{Casas:2017wjh}%
  \BibitemOpen
  \bibfield  {author} {\bibinfo {author} {\bibfnamefont {S.}~\bibnamefont
  {Casas}}, \bibinfo {author} {\bibfnamefont {M.}~\bibnamefont {Pauly}},\ and\
  \bibinfo {author} {\bibfnamefont {J.}~\bibnamefont {Rubio}},\ }\bibfield
  {title} {\bibinfo {title} {{Higgs-dilaton cosmology: An
  inflation\textendash{}dark-energy connection and forecasts for future galaxy
  surveys}},\ }\href {https://doi.org/10.1103/PhysRevD.97.043520} {\bibfield
  {journal} {\bibinfo  {journal} {Phys. Rev. D}\ }\textbf {\bibinfo {volume}
  {97}},\ \bibinfo {pages} {043520} (\bibinfo {year} {2018})},\ \Eprint
  {https://arxiv.org/abs/1712.04956} {arXiv:1712.04956 [astro-ph.CO]}
  \BibitemShut {NoStop}%
\end{thebibliography}%


%
\bibliographystyle{JHEP}

\end{document}